\documentclass[acmlarge]{acmart}

\usepackage{booktabs} 

\usepackage[ruled]{algorithm2e} 
\usepackage [autostyle, english = american]{csquotes}
\MakeOuterQuote{"}
\usepackage{hyperref}
\usepackage{censor}
\usepackage{color}
\usepackage[T1]{fontenc}

\SetAlFnt{\small}
\SetAlCapFnt{\small}
\SetAlCapNameFnt{\small}
\SetAlCapHSkip{0pt}
\IncMargin{-\parindent}

\setcopyright{usgovmixed}

\received{February 2018}
\received{July 2018}
\received[accepted]{October 2018}

\begin{document}
\raggedbottom
\title[SoundSignaling: Realtime, Stylistic Modification of a Personal Music Corpus...]{SoundSignaling: Realtime, Stylistic Modification of a Personal Music Corpus for Information Delivery} 

\author{Ishwarya Ananthabhotla}
\authornote{This is the corresponding author}
\affiliation{%
  \institution{Responsive Environments Group, MIT Media Lab}
  \streetaddress{75 Amherst Street}
  \city{Cambridge}
  \state{MA}
  \postcode{02139}
  \country{USA}}
\email{ishwarya@media.mit.edu}

\author{Joseph A. Paradiso}
\affiliation{%
  \institution{Responsive Environments Group, MIT Media Lab}
  \streetaddress{75 Amherst Street}
  \city{Cambridge}
  \state{MA}
  \postcode{02139}
  \country{USA}}
\email{joep@media.mit.edu}

\begin{abstract}
Drawing inspiration from the notion of cognitive incongruence associated with Stroop's famous experiment, from musical principles, and from the observation that music consumption on an individual basis is becoming increasingly ubiquitous, we present the SoundSignaling system -- a software platform designed to make real-time, stylistically relevant modifications to a personal corpus of music as a means of conveying information or notifications. In this work, we discuss in detail the system's technical implementation and its motivation from a musical perspective, and validate these design choices through a crowd-sourced signal identification experiment consisting of 200 independent tasks performed by 50 online participants.  We then qualitatively discuss the potential implications of such a system from the standpoint of switch cost, cognitive load, and listening behavior by considering the anecdotal outcomes of a small-scale, in-the-wild experiment consisting of over 180 hours of usage from 6 participants. Through this work, we suggest a re-evaluation of the age-old paradigm of binary audio notifications in favor of a system designed to operate upon the relatively unexplored medium of a user's musical preferences.
\end{abstract}

%
%


\setcopyright{acmcopyright}
\acmJournal{IMWUT}
\acmYear{2018} \acmVolume{2} \acmNumber{4} \acmArticle{154} \acmMonth{12} \acmPrice{15.00}\acmDOI{10.1145/3287032}

%
%



\maketitle

\renewcommand{\shortauthors}{I. Ananthabhotla et al.}

\section{Introduction}

\subsection{Background}

\textit{Addicted to Distraction}, reads the title of a New York Times front-page Sunday review article.  \textit{Brain, Interrupted}, reads yet another, published a few years earlier. Both titles are a testament to the power that technological distractions have wielded over our society at large, as portrayed by both public media \cite{NYT1,NYT2} and scientific assessment \cite{adamczyk2004if,horvitz2001notification,czerwinski2000instant,czerwinski2004diary}. Statistics show that the average American checks his or her phone on the order of 100 to 200 times a day, for a mean duration of approximately 1 minute \cite{checkphones, bohmer2011falling}; that an individual is likely to receive hundreds of notifications a day \cite{pielot2014situ, mehrotra2015designing}, and spend on average six hours a day interacting with his or her email \cite{adobesurvey}.

The distractions stemming from the technology we use on a daily basis, and notifications in particular, have been demonstrated to be detrimental from a task productivity and attentional standpoint.  Work done by \cite{adamczyk2004if,horvitz2001notification,czerwinski2000instant,czerwinski2004diary} has explicitly shown that notifications are a source of interruptions, negatively impacting task execution, performance, and memory \cite{monk2002attentional}.  Though notifications have been shown to cause stress and frustration independant of the context of the work which they interrupt \cite{mark2008cost}, it has also been established that simply "unplugging" from technology is not, at least in the short term, a beneficial antidote \cite{oulasvirta2012habits}.  Users acknowledge that notifications are disruptive, but continue to enable them as a means of remaining aware of activity and communication within their operating contexts \cite{iqbal2010notifications}.  Additionally, users value control over their ability to respond to notifications, yet tend to respond to them immediately and more often than they realize, as shown by self-assessments \cite{iqbal2007disruption}. 

The doubled-edged sword that is the nature of notification design has and continues to generate a substantial amount of research in the field of HCI -- work by \cite{turner2015interruptibility, okoshi2015reducing, chen2004using, ho2005using, fischer2011investigating, pejovic2014interruptme} is only a small selection.  In particular, two major themes from this body of literature served as an inspiration for the investigation in this work:

\textit{\textbf{Switch Cost.}} Monsell, in a canonical work from the field of cognitive science, demonstrated early on that switching from one continuous task to another, when prompted to do so by an involuntary stimuli, necessarily results in quantifiable loss in performance on the task being switched to \cite{monsell2003task}.  This was further corroborated in uncontrolled contexts, such as in the workplace, where it was found that added stress, speed, and a difficulty in returning to an original task were all a part of the "switch cost" associated with commonplace interruptions \cite{mark2008cost, czerwinski2004diary}.

\textit{\textbf{Independence from Cognitive Load.}} HCI studies have also shown that the mental load associated with the current task at hand plays a signification role in determining the level of disruption perceived by a user when a notification is received \cite{mehrotra2016my}. Yet, the notifications that we receive today are entirely \textit{binary} -- the "noticeability" of a notification is usually not a function of one's level of mental alertness or mental processing.

In the interest of investigating other modalities that may allow for a further exploration of the challenges presented, we consider a novel means to information delivery that capitalizes on a relatively unexplored but ubiquitous medium -- a user's personal music collection.

\subsection{Our Approach}

In a world where access to digital music is widespread, there is no denying the fact that individuals across the globe consume music at a rate greater than ever before.  A recent study done in the UK demonstrated that, at the current rate of music consumption, the average person will have listened to music for 13 years out of their lifetime; similar figures in the US show that the average American listens to music for approximately 40 hours per week \cite{GMR2017,ukstats}.

Research also suggests that the music we listen to is often music that we know well \cite{spotdiv}. Long term research has shown that repetition is an integral part of the listening experience, and according to one famous compendium in the field of music cognition, "99 percent of all listening experiences involve listening to musical passages that the listener has heard before." \cite{margulis2014repeat}

In addition to these statistics, we draw inspiration from one of the most well-known works in experimental psychology on \textit{cognitive incongruence}, demonstrated by the famous Stroop experiment \cite{stroop1935studies}.  In his work, Stroop demonstrates that the dual processing of language and color elicits "race conditions" in processing and attention, resulting in confusion that causes a derailment of the identification task \cite{johnson2004attention}.  It has also been demonstrated that this "confusion" is modulated by an individual's cognitive load \cite{chen2003attentional}. Here, we seek to extrapolate these results from visual to auditory processing.  We intuit that, for many who listen to music regularly while completing an additional task, the stimuli are jointly perceived to be congruent. However, subtle, musically relevant but distinctly artificial modifications made to this music may be perceived to be incongruent (at a conscious or sub-conscious level), perhaps eliciting an attentional shift that may be less pronounced than when presented with a standard notification that surfaces through music already playing, or is perhaps naturally a function of cognitive load and the task at hand.

Building upon this information, we present the SoundSignaling system, a software platform that uses an individual's personal corpus of music as a canvas for subtle, real-time audio modifications and manipulations to convey "notifications" to a user.  These modifications, which may be driven by any notification source of interest, are designed to be stylistically relevant to the audio being operated on as a result of an online assessment of the track's genre and low-level musical features, but agnostic to the track itself, allowing for flexibility, large-scale use, and most importantly, customization to an individual's taste in music.  Moreover, these audio modifications are made at three "levels of subtlety" per genre classification, with an increasing likelihood of detection. 

Specifically, the contributions made in this paper are as follows:
\begin{enumerate}
  \item A detailed presentation of the design and technical implementation of the SoundSignaling software application, including a motivation of the design choices from a musical perspective.
  \item The goals, procedures, and outcomes of a crowd-sourced evaluation designed to validate the set of algorithms chosen for genre-specific musical modifications.  Through an online experiment consisting of approximately 200 identification tasks, we show that the operating assumptions of being more likely to perceive notifications when they are made in familiar as opposed to unfamiliar music, and at higher "levels of subtlety", are valid.
  \item An anecdotal discussion of outcomes from a small-scale, in-the-wild evaluation designed to understand patterns in user interaction with the system that took place over a period of 10 days consisting of 180 hours of collective usage, 67 independent listening sessions, and 157 email notifications from 6 participants randomly selected from our university.  
\end{enumerate}

\section{Related Work: Music as a Medium}

In recent years, the concept of ambient notifications, or the idea of embedding information into visual, auditory, or tactile stimuli that are \textit{already} present in the environment has been explored in a variety of manifestations \cite{wiehr2016challenges, matviienko2015towards, gellersen1999ambient, ishii1998ambientroom}. Regarding the element of sound, the task of conveying information using audio as a medium has typically been done via sonification, which entails developing a custom mapping between some properties of the information and generative properties of the audio \cite{dubus2013systematic}. While sonification approaches have typically been studied with the intent of allowing for more detailed, high-resolution exploration of data than visualizations can provide, these approaches are typically facilitated by custom mappings motivated by specific data sources -- and hence users require time and training within a particular context to effectively interpret the information being conveyed \cite{kramer2010sonification}.  

Focusing specifically on alternative auditory notifications, previous work by \cite{jung2008ambience, jung2010non, butz2005seamless} developed custom recorded, person-specific audio notifications intended to fit seamlessly into an ambient musical composition.  This was done with the intent of making publicly audible sound notifications a private affair, as only the recipient of the notification would be aware of the meaning of the embedded audio cue.  While an important foray into the space of musical manipulation for information delivery, we recognize that the work is not generalizable to individuals and their personal collections of music (a more likely scenario for music consumption than ambience), and require explicit training in order to enable recognition of the auditory cues.

Finally, work by \cite{barrington2006ambient} develops a system to add "musical effects" (such as reverb and low-pass filtering) to an audio track as a sonification approach, driven by facial affect sensor data.  However, the choice of "effects" are somewhat arbitrary; the modifications made to the music are not relevant to the music itself.  To the best of our knowledge, our work is novel in this regard, and but is also a novel approach to the idea of sonification itself -- rather than learnable perturbations whose behavior is governed by a mapping between data and pre-determined audio, our approach explores musically relevant but non-patternizable, dynamic manipulations that operate on a personal music collection as a more familiar auditory medium.

\section{Technical Implementation}

\subsection{Overview and Real-time System Design}

\begin{figure}
  \includegraphics[width=0.8\textwidth]{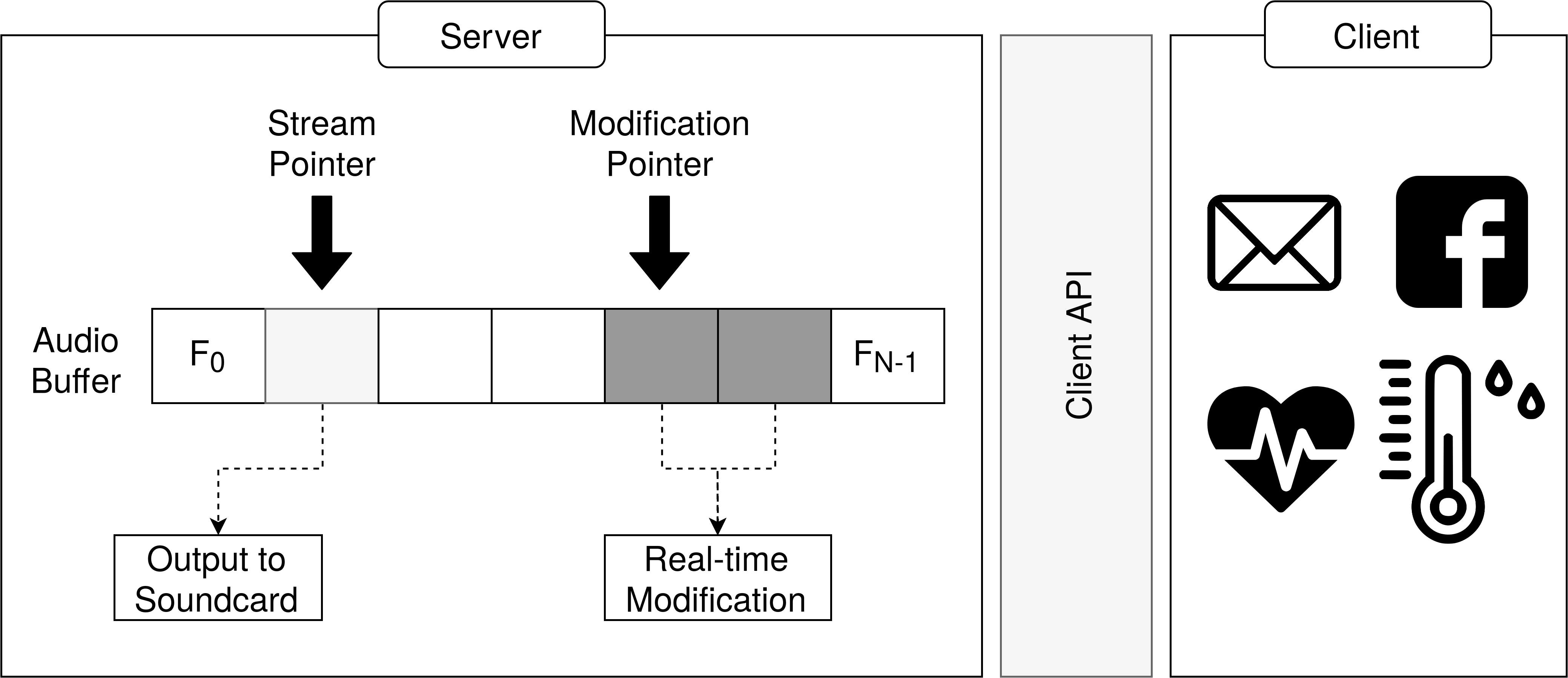}
  \caption{An illustration of the client-server model governing the SoundSignaling system.}
  \label{server-client}
\end{figure}

The software presented in this work has a general architecture that is given in Figure \ref{server-client}, and follows a simple client-server model to demonstrate utility across a wide range of applications.  A user first uploads audio selected for listening into the application, and supplies metadata pertaining to the audio (the genre and/ or the time signature) to the extent possible. The audio playlist is then pre-processed by the application "server" (discussed in greater detail below).  After completion, the server begins a process to load audio frames into a buffer, from which frames are read by a secondary process and output to the system's sound card.  However, the server is also engaged with an external "client" application providing stimuli to the server using the Client API.  While the stimuli may be triggered by any external process a user may wish to monitor (a fluctuating home thermostat, or Facebook account notifications, for example), the Client API allows for a time-stamped request for a genre-specific music modification, with an associated "subtlety level", to be handed to the server.  In this work, we refer to the "subtlety level" as our hypothesized likelihood of perception, or metric of obviousness, assigned to a given type of genre-specific modification. The server, in turn, manipulates the audio data buffer several frames ahead of the streaming process' current pointer, such that the user observes a perturbation in the original music as a real-time reflection of the external event.

The algorithms governing the nature of the modification as a function of the music's characteristics are detailed in the sections below.  Additionally, samples of audio in each genre category with modifications made at each level can be heard at \href{http://resenv.media.mit.edu/soundsignaling}{resenv.media.mit.edu/soundsignaling}, to better illustrate the implementation details presented below.

\subsection{Genre Specific Modification}

For the purpose of this work, four classes of modifications were designed pertaining to the following broad musical genres -- Classical, Jazz, Blues, and Popular (Pop).  These genres were intentionally chosen as base categories for their distinct characteristics, allowing the authors to engineer modifications explicitly pertinent to these characteristics from a musician's perspective.  For example, Classical music is generally characterized by dynamically varying tempo, amplitude, and depth of orchestration, suggesting that perturbing these parameters within a piece would present the desired incongruence to a listener familiar with the genre or the piece in particular, but would not seem jarring or entirely out of place.  Needless to say, it is quite likely that a user wishes to use the system with a playlist comprised of genres other than the four mentioned.  In this case, a user has two options when entering metadata into the system about his or her playlist: (1) a user may choose from the genre keywords shown in Table 1, which internally map to the four categories of modifications based on similar musical characteristics (as are relevant to the modification algorithms), or (2) in the case where the genre label is unknown, a user may simply leave this parameter blank, allowing the system to automatically determine the most appropriate class of modifications to perform as a function of the nature of the audio (see Section \ref{autosort} and Figure \ref{autosort_figure} for further details).

\begin{table}%
\caption{This table shows the list of possible genre keywords that can be used to label audio tracks supplied by a user, and the internal mapping between these keywords and the four categories of modifications.}
\label{keywords}
\begin{minipage}{\columnwidth}
\begin{center}
\begin{tabular}{ll}
  \toprule
  \textbf{Genre Category}  & \textbf{Keywords}\\
  Classical       & classical, rhythmless-instrumental, choir, avant-garde, soundtrack\\
  Blues           & blues, rock, hip-hop, R\&B, soul, strong-rhythmic, disco, rap\\
  Jazz            & jazz, rhythmic-instrumental, electronic, easy-listening\\
  Pop             & pop,country, folk, latin, gospel\\
  \bottomrule
\end{tabular}
\end{center}
\end{minipage}
\end{table}%

\begin{figure}
  \includegraphics[width=1.0\textwidth]{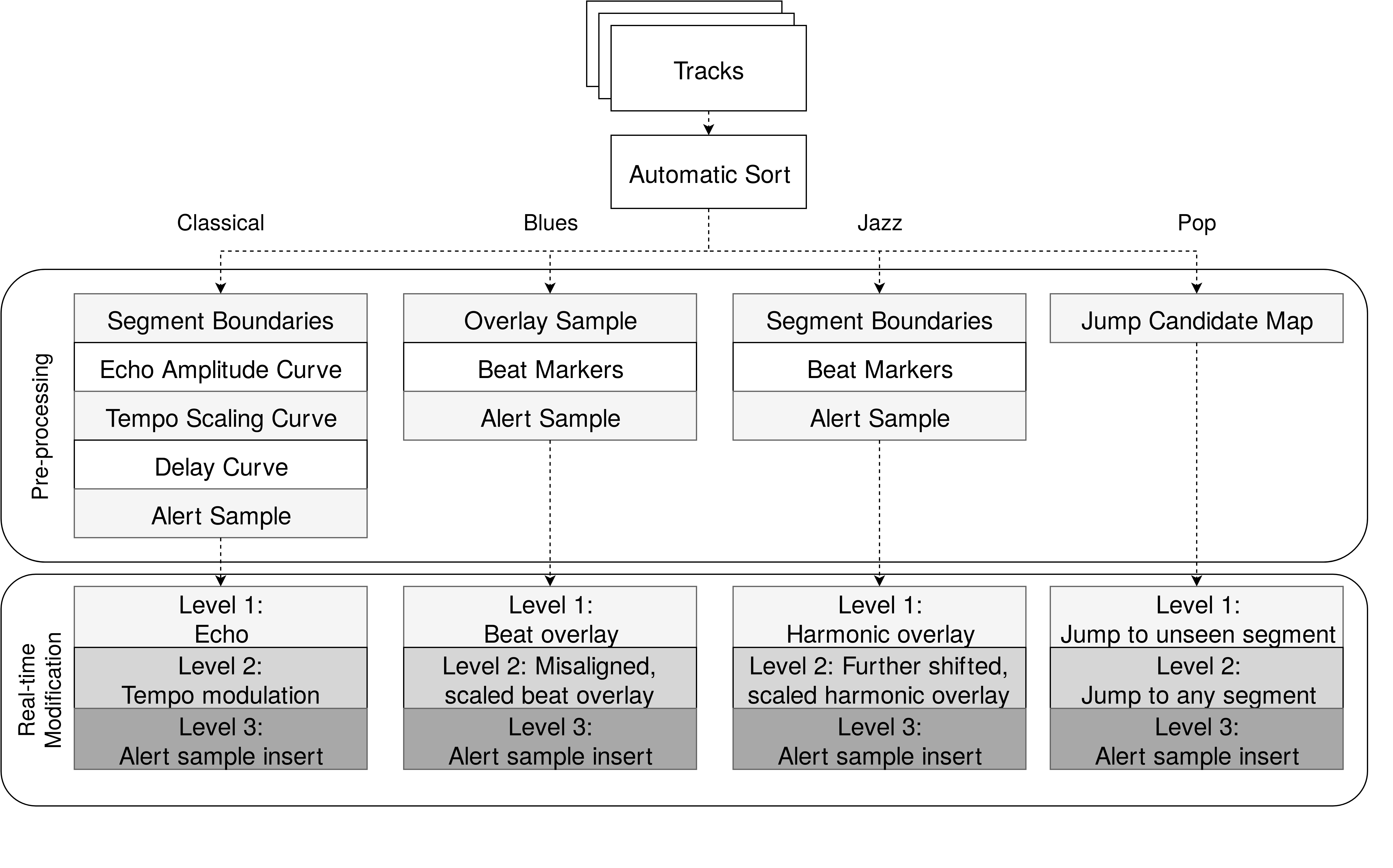}
  \caption{An overview of the pre-processing and real-time modification algorithms used in the SoundSignaling system.}
  \label{overview}
\end{figure}

A detailed breakdown of the algorithms used to achieve the modifications described in this section is shown in Figure \ref{overview} and will be discussed below.  While the modification objective of each algorithm is the result of the authors' musical intuition, the processing phases of each build upon principles typically used in the Music Information Retrival (MIR) communities for audio analysis or feature extraction.

\subsubsection{"Classical" Algorithms}

\begin{figure}
  \centering
  \includegraphics[width=0.75\textwidth]{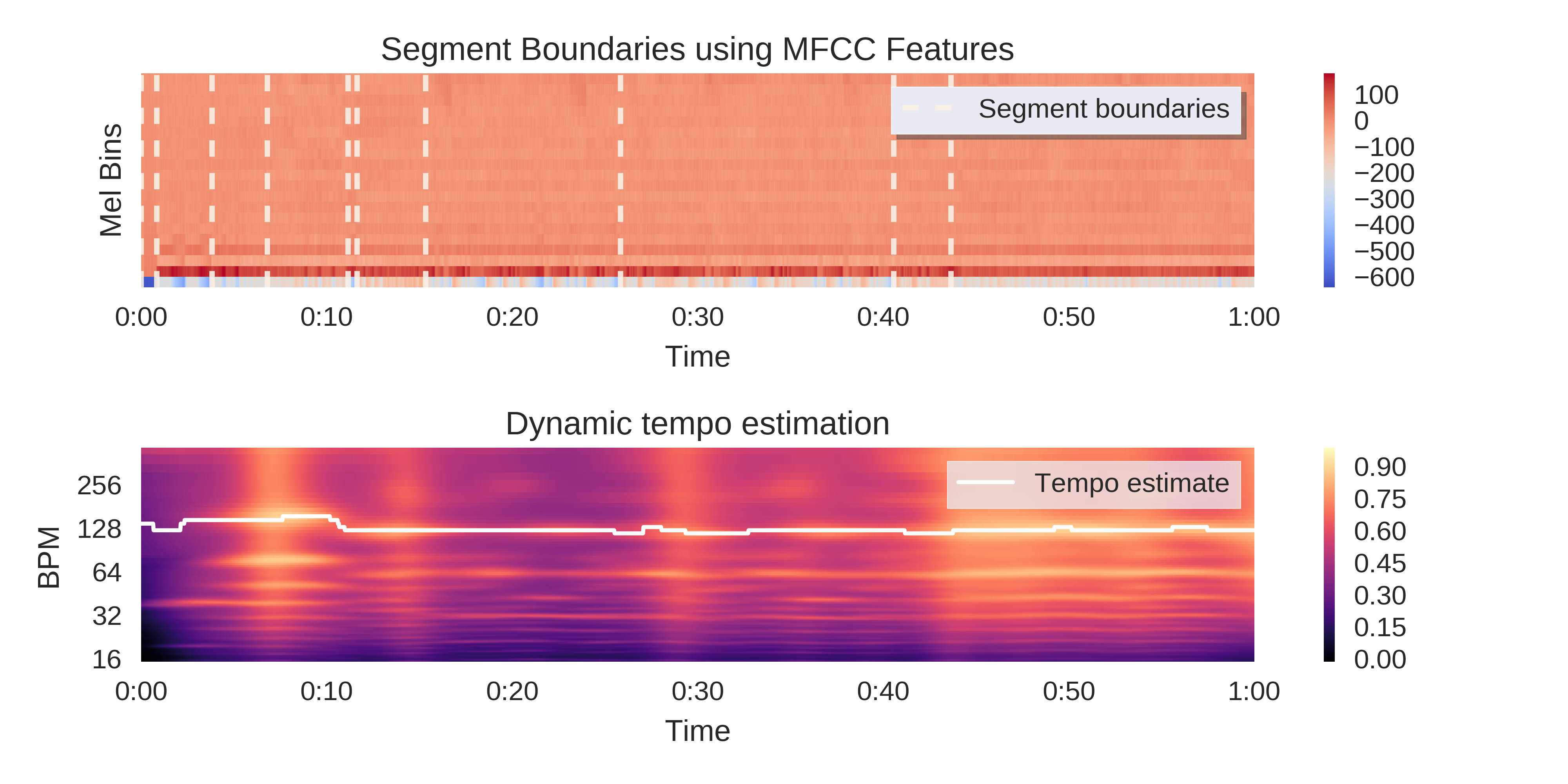}
  \caption{Top: MFCC feature extraction and segmentation on a sample classical track; Bottom: Dynamic tempo estimation on the same track.}
  \label{classical1}
\end{figure}

\begin{figure}
  \includegraphics[width=0.45\textwidth]{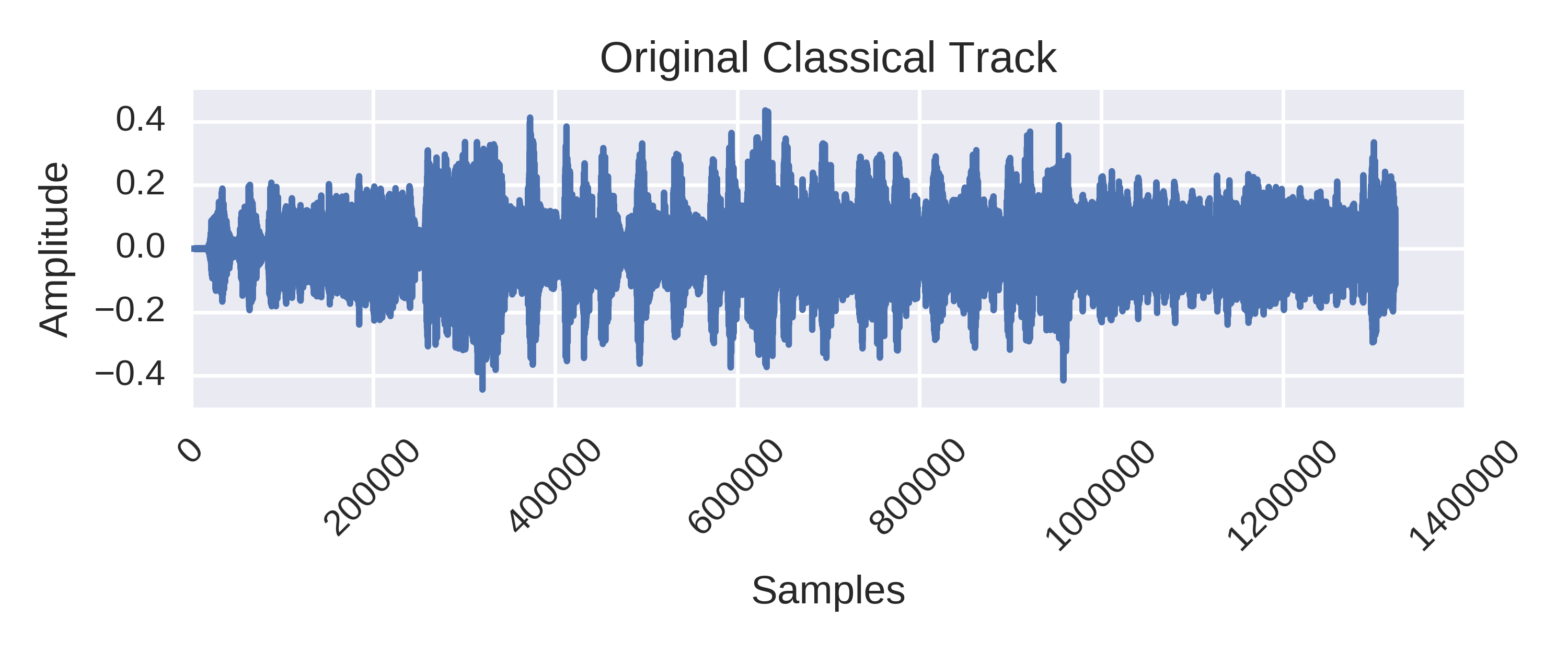}
  \includegraphics[width=0.45\textwidth]{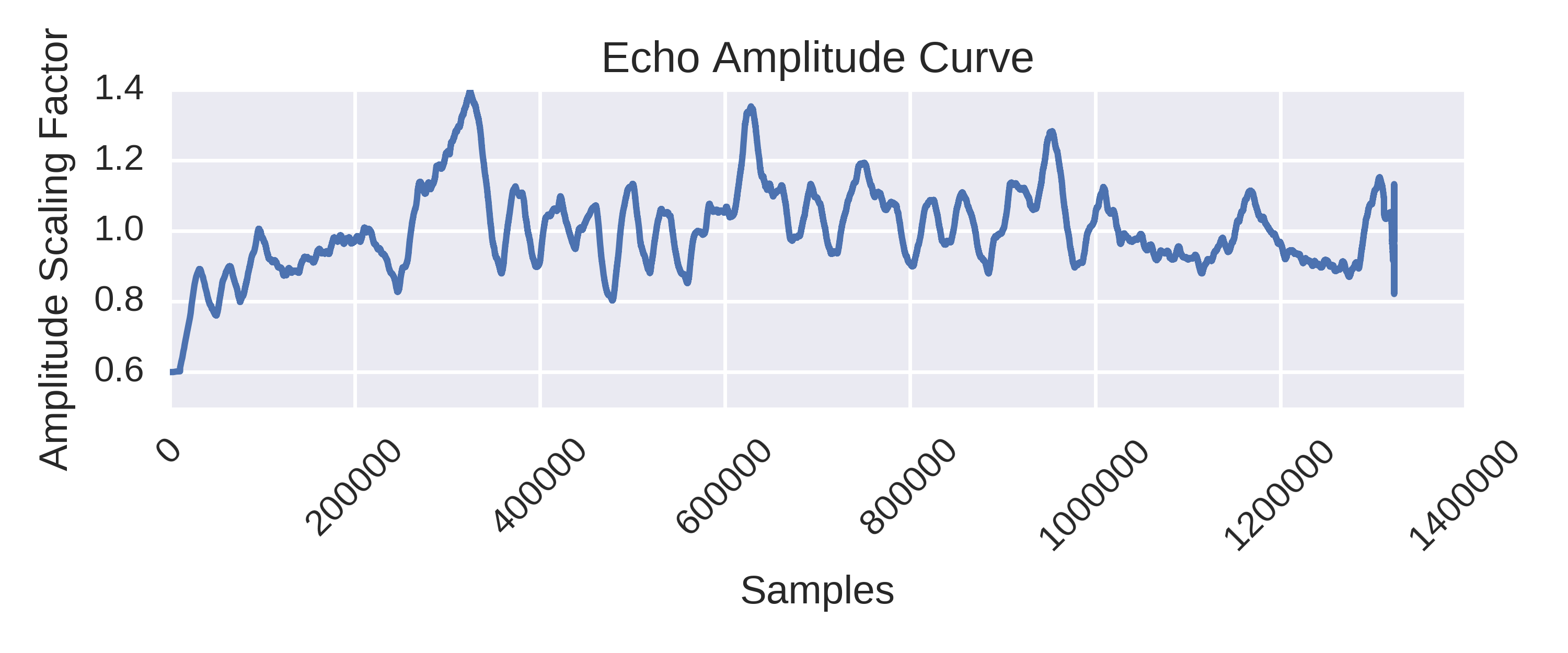}
  \includegraphics[width=0.45\textwidth]{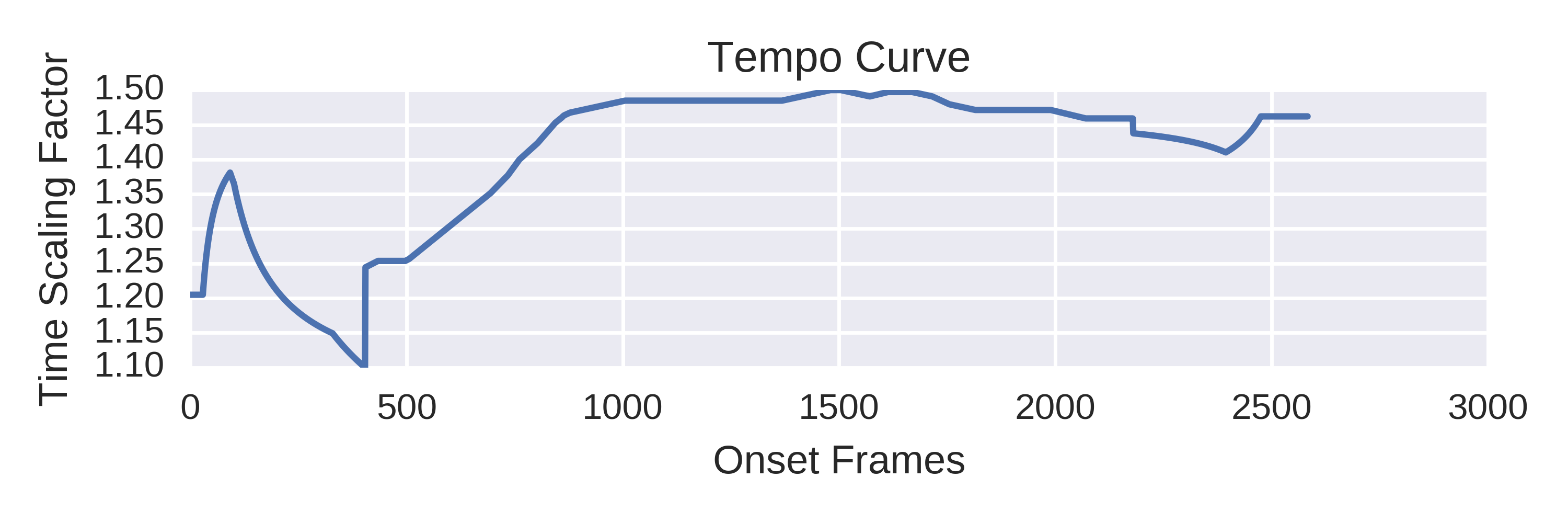}
  \includegraphics[width=0.45\textwidth]{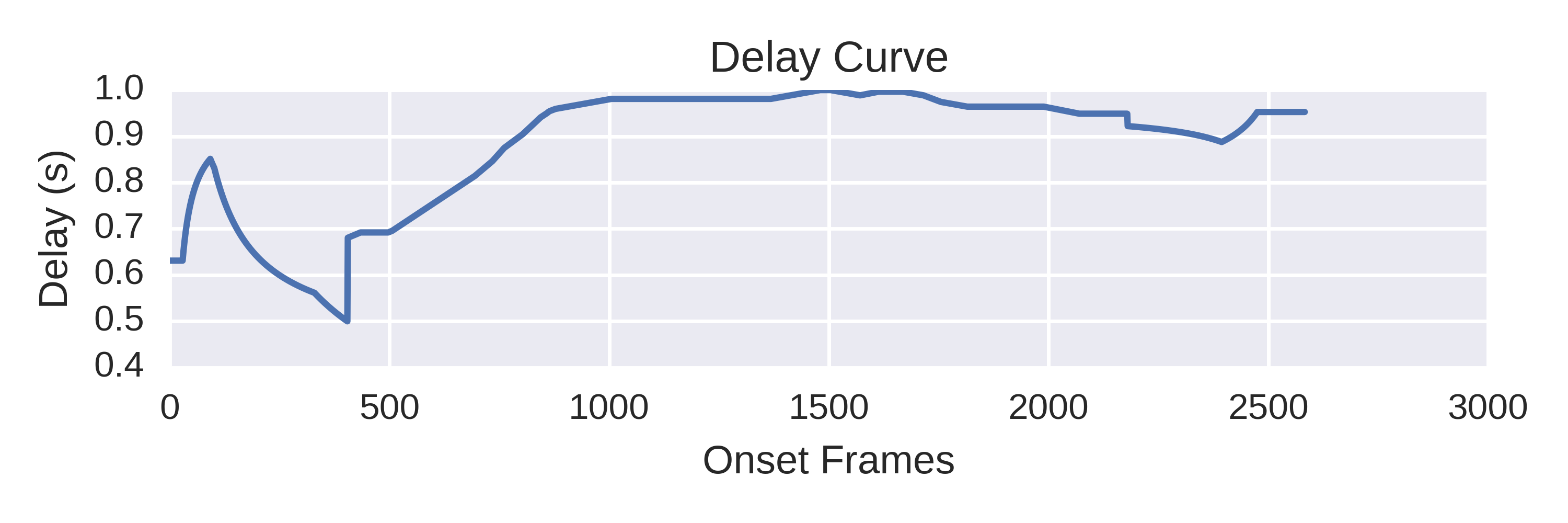}
  \caption{An illustration of the echo amplitude, tempo, and delay curve extraction process.}
  \label{classical2}
\end{figure}

As mentioned in the example, classical music is characterized by dynamic pace, volume, and orchestration.  The modification process for user music categorized as classical capitalizes on this observation at three levels of subtlety.  At the most subtle level (here on refered to as Level 1), the modification process introduces an echo to the audio stream at the start of the next nearest musical section, which may be characterized by a change in motif, instrumentation, tempo, timbre, etc. In this context, an echo is defined by the re-addition of an existing audio sample at a later point in time, and is parameterized by its delay and amplitude in relation to the original. To determine these parameters, the preprocessing stage begins by extracting a given track's MFCC coefficients and performs agglomerative clustering to determine sectional bounds (see Figure \ref{classical1}) \cite{turnbull2007supervised, klapuri2010audio}.  It then computes the track's onset strength envelope to obtain a dynamic tempo curve, which is smoothed by a low-pass filter to obtain a low-fidelity representation of the entire track's tempo as a function of time \cite{alonso2004tempo}. This is used to compute a "delay curve", a metric of how much an audio sample should be delayed as a function of the track's current tempo in time. Similarly, an "echo amplitude curve", or a metric of the re-inserted audio sample's amplitude as a function of the amplitude of the audio in its neighborhood, is computed from an amplitude envelope of the audio track (see Figure \ref{classical2}).
The computation of the delay curve $D(t)$ as a function of the track's tempo estimate $T(t)$ is given as below:
\begin{equation}
D(t) = D_{max} + \frac{(D_{min} - D_{max})}{(T_{max} - T_{min})} * (T(t) - T_{min})
\end{equation}
with upper and lower scaling bounds of $D_{max}$ and $D_{min}$ and tempo bounds of $T_{max}$ and $T_{min}$.

Similarly the computation of the echo amplitude curve $E(t)$ as a function of the track's low pass filtered amplitude curve $A(t)$ is given by: 
\begin{equation}
E(t) = E_{max} + \frac{(E_{min} - E_{max})}{(A_{max} - A_{min})} * (A(t) - A_{min})
\end{equation}
with upper and lower scaling bounds of $D_{max}$ and $D_{min}$, and maximum and minimum amplitude curve values of $A_{max}$ and $A_{min}$. The segment bounds, delay curve, and echo amplitude curve are stored by the proprocessing stage for later use. When a Level 1 request is delivered to the server actively streaming a track tagged as "Classical", this information is used to select a segment in the audio buffer of a fixed duration (typically 1-2 seconds) beginning at the next nearest segment bound, determine its delay and amplitude, apply a hanning window to mitigate edge artifacts, and reinsert it by superimposition ahead of the stream pointer in order for the modification to be output to the user in realtime.

At the next level of subtlety (referred to as Level 2), the modification process increases the tempo of a fixed-length passage of the audio before it is transitioned back to its original tempo.  The rate at which the tempo is scaled is determined from the low-fidelity tempo curve, already computed and stored by the preprocessing stage as mentioned above (see Figure \ref{classical2}).  The computation of the tempo scaling curve $R(t)$ as a function of the tempo curve $T(t)$ is given as:
\begin{equation}
R(t) = R_{max} + \frac{(R_{min} - R_{max})}{(T_{max} - T_{min})} * (T(t) - T_{min})
\end{equation}
with upper and lower scaling bounds of $R_{max}$ and $R_{min}$ and tempo bounds of $T_{max}$ and $T_{min}$. Upon presentation of a client request, the server application uses the pre-computed bound information to identify a logical start for the segment, computes the new tempo, and using a phase vocoder-based tempo scaling algorithm (so as to preserve pitch) \cite{ellis2002phase}, processes the selected audio segment to create a replacement for the sample located in the audio buffer. Once replaced, the remainder of the samples ahead of the segment are simply advanced to account for the loss in content. 

At the level of least subtlety (Level 3), a portion of the audio from a different location in the buffer is resampled and inserted at the nearest segment bound.  This was designed to closely mimic the concept of a standard audio notification or icon to ensure perceptibility, but in such a fashion that still ensures a stylistic connection between this "alert" the original audio track.  To achieve this, a technique for representative selection extraction developed by the authors in a previous work is used to identify an audio selection for this "alert" on the basis of length, homegeneity, and monophony from the original audio track, as a part of the preprocess stage \cite{ishwarya}.  In realtime, the server application simply loads this pre-chosen segment, applies a tapered window to mitigate edge effects after trimming it to the desired length (a fixed parameter, usually 1 second), and inserts it at the start of the next segment.  The set of samples in the audio buffer that overlap with the "alert" selection are then delayed to account for the additional content. 

The modification process at a Level 3 follows the same procedure across all genre categories to maintain consistency at what is intended to be the most perceivable level of subtlety; as such, a description of the Level 3 algorithms for the genre categories to follow will be left out to eliminate redundancy.

A visual representation of the preprocessing techniques employed by the algorithms in this genre category is given in Figure \ref{classical1} and Figure \ref{classical2}.

\subsubsection{"Blues" Algorithms}

\begin{figure}
  \includegraphics[width=0.65\textwidth]{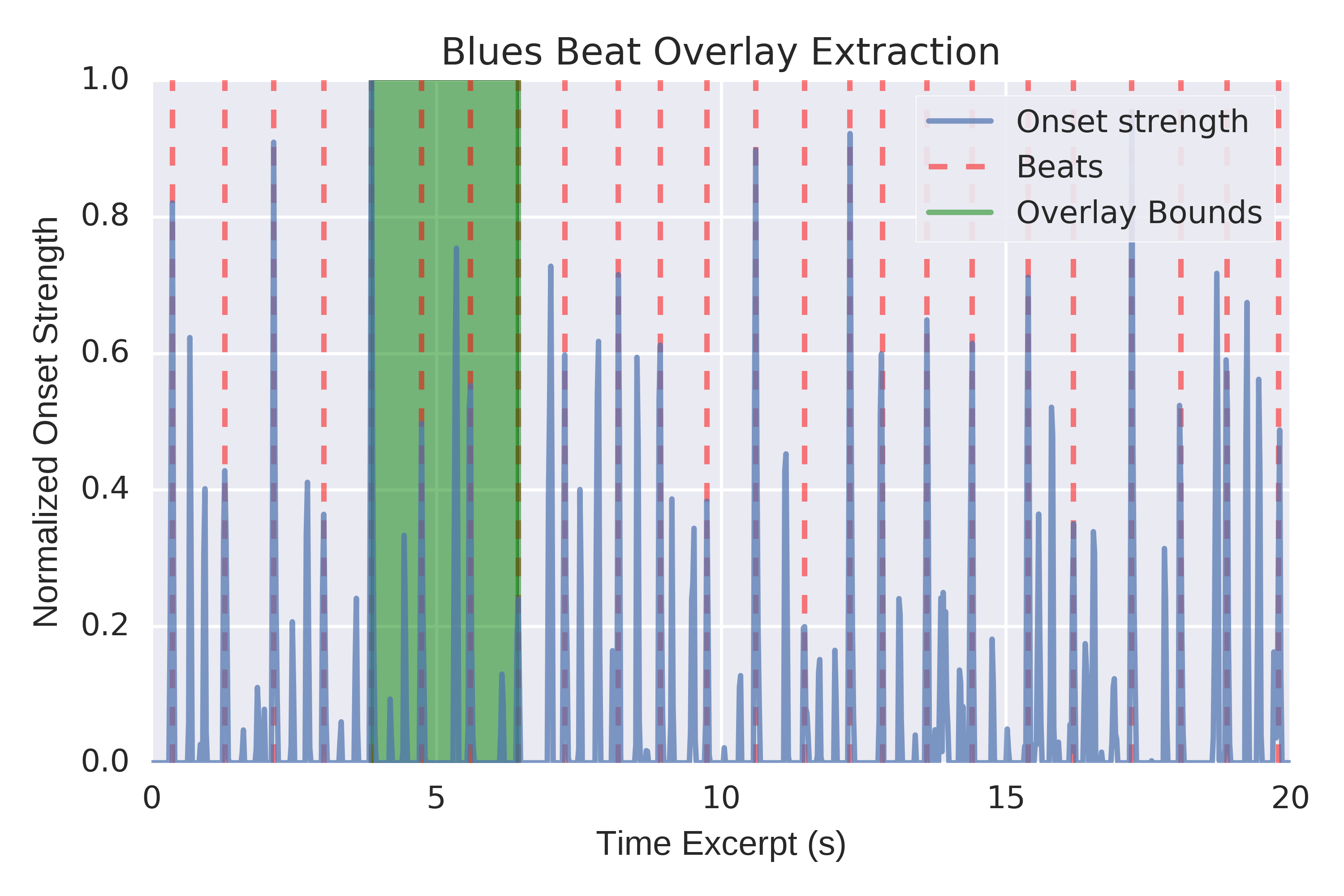}
  \caption{A demonstration of the percussive track onset strength calculation and beat overlay sample selection from a sample blues track.}
  \label{bluesoverlay}
\end{figure}

While Blues as a musical genre has many flavors and forms, it is well known for its standard, repetitive structure.  As a result, we determined that Blues music would be a good candidate for resampling of rhythmic units, as would be any genre typically containing palpable, steady rhythm tracks.  For any audio track tagged as "Blues", the Level 1 modification entailed the selection of a prominent rhythmic passage from one section of the track overlaid onto the percussive track in another section of the audio.  At a Level 1 modification, this overlay was applied at an amplitude comparable to the existing percussive track, and aligned with pre-computed beat markers.  At a Level 2, the overlay was applied at an upscaled amplitude, and in slight misalignment with the rhythm, so as to make the incongruence increasingly apparent.  In order to achieve this, we first use a dynamic time-warping based beat-tracking algorithm \cite{ellis2007beat} for the entire audio track in the preprocessing stage, and store the timestamps of all determined beats for later use (see Figure \ref{bluesoverlay}). Additionally, the audio track is decomposed into its harmonic and percussive components \cite{fitzgerald2010harmonic}, and the onset strength of the percussive component is calculated to locate passages with strong rhythmic elements.  Several beats from within such a region are selected to meet a fixed duration (typically 1-2 seconds) and are grouped together as an overlay segment, which is also saved in the preprocessing stage.  In realtime, the server application simply determines the location of the next nearest beat ahead of the stream pointer in the audio buffer, applies a tapered window to the overlay segment, shifts and/ or scales the segment as determined by the subtlety level, and superimposes the segment upon the buffer.

A visual representation of the preprocessing techniques employed by the algorithms in this genre category is given in Figure \ref{bluesoverlay}.

\subsubsection{"Jazz" Algorithms}

Jazz music, while arguably one of the most difficult genres to define, is most noteworthy for its element of improvisation.  When coupled with "swing" and "blue notes", the improvisational component often results in atypical polyphony or progressions, making this genre category an excellent candidate for re-harmonization as a means for presenting incongruence.  At the preprocessing stage, segment boundaries are drawn using MFCC coefficient clusters (as described above) and are stored for later use.  In realtime, when a Level 1 or Level 2 request is sent to the server application, a fixed-length sample (typically 1-2 seconds in length) beginning at the next nearest segment boundary is selected and pitch shifted away from the tonic, to what is likely to be perceived as dischordant by most listeners\footnote{This is, of course, non-deterministic without knowing the key signature of the piece or the standard.} (for example, 3 semitones).  A sample being modified under a Level 2 request is pitch shifted further away from the tonic than under a Level 1 request, and is additionally amplified beyond the volume of the original segment, before it superimposed onto the appropriate location in the audio buffer ahead of the stream pointer.

\subsubsection{"Pop" Algorithms}

Pop music is often associated with audio tracks that are structurally and musically redundant, with chorus lines that appear after every stanza or verses that have identical tunes, for example.  For this type of music, we use a set of algorithms that seek to capitalize on this redundacy. Here, we implement a process that is inspired by and is very similar to the design of the Infinite Jukebox, a web platform that extends an audio track infinitely by jumping between similar sounding musical segments that correspond to the beat of the track \cite{lamere2012infinite}. At both a Level 1 and Level 2 modification, the server application redirects the stream pointer to an alternate, musically similar location in the audio buffer, at the start of the next nearest beat.  As a part of the preprocessing stage, we compute a mapping between every beat in the track and possible candidates that can be "jumped" to from that beat, using an adaptation of the method in \cite{lamere2012infinite}. In realtime, when a Level 1 request is handed to the server application, a jump location is chosen at random from the set of candidates associated with the next nearest beat that have not been visited before.  If a Level 2 request is handed to the server application instead, a jump location is chosen specifically from the set of candidates that have been visited before, to increase the likelihood of perception by the user.  

\subsubsection{Automatic Sort}\label{autosort}

\begin{figure}
  \includegraphics[width=0.32\textwidth]{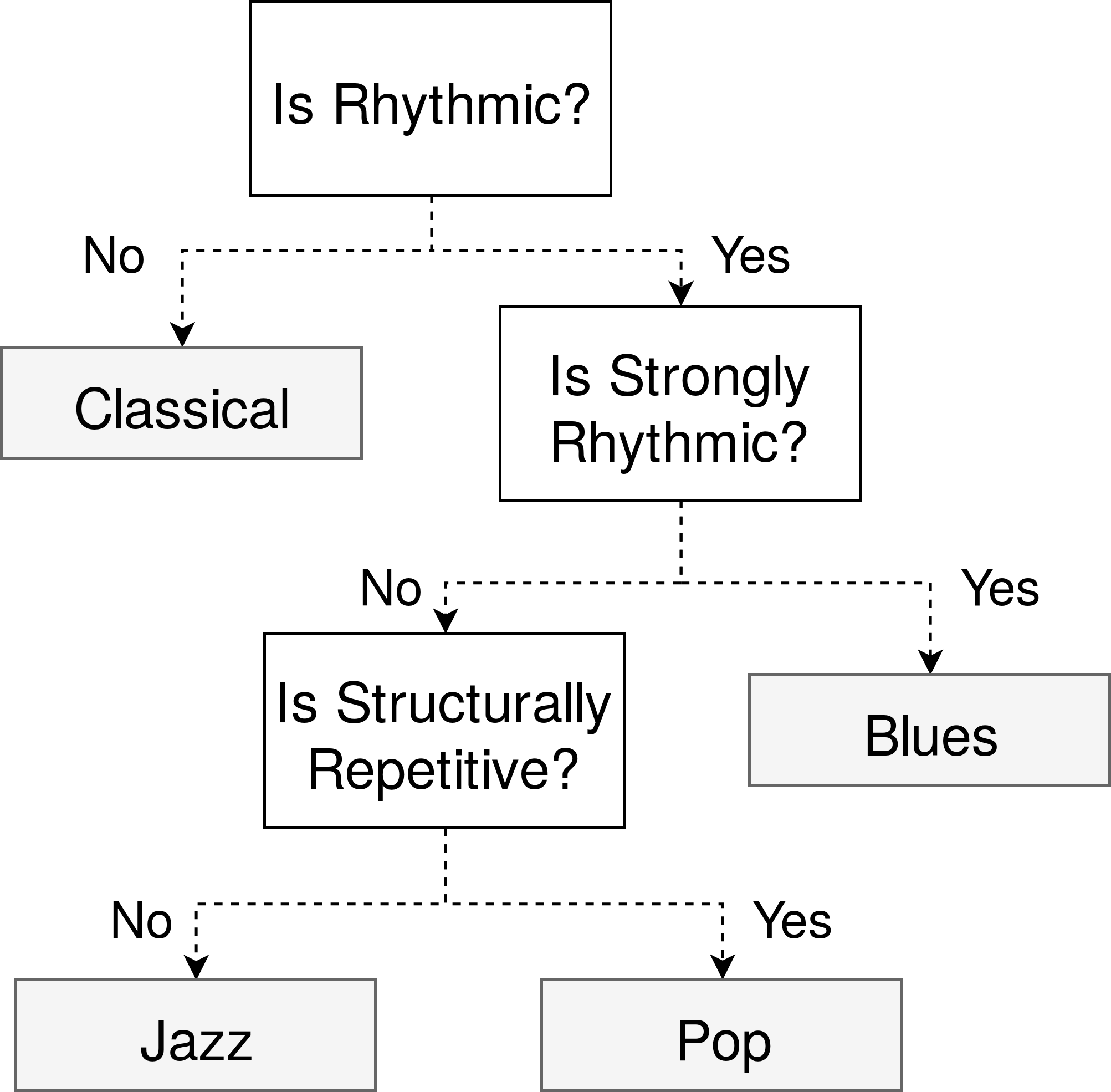}
  \caption{Decision tree governing the automatic category sorting functionality of the SoundSignaling platform.}
  \label{autosort_figure}
\end{figure}

As discussed earlier, users are additionally given the option to leave information out of the genre field when supplying metadata for the tracks they wish to upload to the system, in case the genre is unknown to the user. In this case, the application attempts to automatically place the tracks into one of the four modification classes by first processing the audio.  It is important to note that automatic genre classification from audio signals is an active area of research in the MIR community, but such deep-learning based state-of-the-art models are beyond the scope of this work.  The simple classification method presented here is designed to analyze and sort audio tracks based \textit{only} on properties that are relevant to the types of modifications being performed.

As shown in the decision tree in Figure \ref{autosort_figure}, every unlabeled track is first assessed for its rhythmic content.  To do this, the onset strength of every beat in the track's percussive component is compared against two thresholds, $r_{1}$ and $r_{2}$, to determine the number of strong and the number of extremely strong beats. These values are in turn evaluated against two success percentages, $s_{1}$ and $s_{2}$, identifying as a result whether the audio track is considered rhythmic, strongly rhythmic, or neither.  Each track is then assessed to determine the degree to which it is structurally repetitive by computing the total number of unique possible "jump" candidates across all beats, and evaluating this value against a threshold $p_{1}$.\footnote{The mechanism used to determine all thresholds in this section, a manual labeling and clustering process, is detailed in a previous work \cite{ishwarya} by the same authors.} A walkthrough of the decision tree using the boolean values resulting from the above comparisons produced the genre categorization that was adopted by the remainder of the application pipeline.

\section{Crowd-sourced Evaluation: Concept Validation}

\begin{figure}
  \fbox{\includegraphics[width=0.75\textwidth]{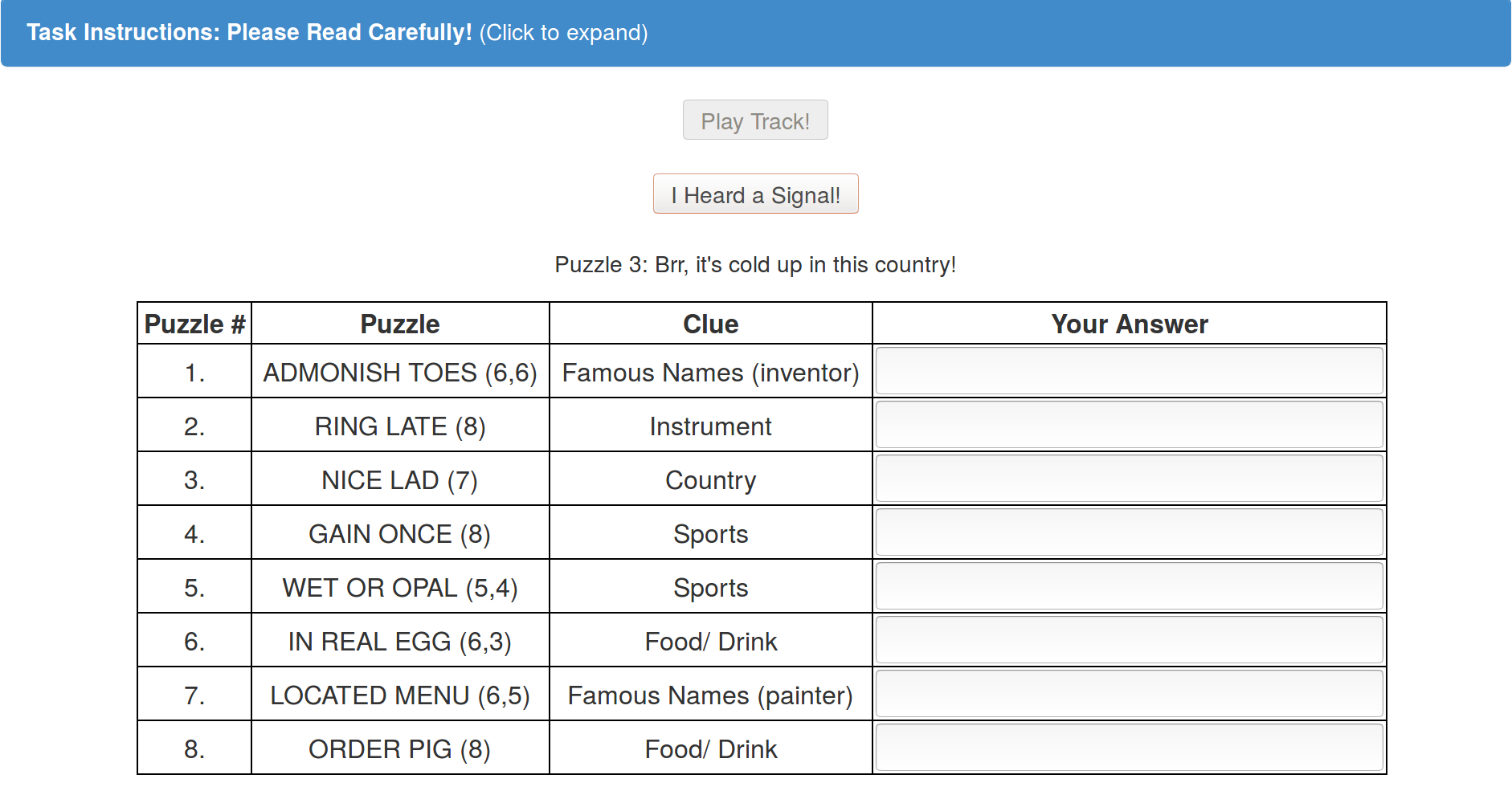}}
  \caption{Screenshot of the interface shown to participants as one task of the crowd-sourced study.}
  \label{screenshot}
\end{figure}

In an experimental setting, we sought to assess the validity of the baseline assumptions made in designing the SoundSignaling system and whether the algorithmic choices described in the sections above achieved the intended behavior of the system.  More specifically, we sought to gain insight into the following: whether or not the chosen algorithms effectively modified the music in a such a way that individuals familiar with a particular track or musical genre were more likely to perceive modifications than those who were not; whether, in the context of a user being presented with a highly familiar track, the three stages of music modifications per genre corresponded to increasing "levels of obviousness", in that a modification was more likely to be perceived if performed at a higher level; and finally, whether the aforementioned behaviour could be demonstrated in a context where the user is occupied by some unrelated mental task, as might be the case when such a system is used in the wild.

\subsection{Study Procedure}
To investigate these key questions, we conducted an online, crowd-sourced evaluation using Amazon's Mechanical Turk platform, consisting of 10 independant 5 minute tasks.  A single online worker could, but was not required to, complete all 10 of the tasks, but was barred from repeating any task a second time.  Each worker was required to have a "Master" status (a mark of recognition of high work quality) in order to participate in the study, and received \$0.50 as compensation for a single task, which was selected at random from the available pool.  The results presented are comprised of 200 such tasks completed by 50 independent turk workers.

In each task, workers were presented with the interface shown in Figure \ref{screenshot}. An instruction window (minimized in the view shown) explained that, when ready, users were required to click the "Play!" button to trigger audio playback and would not be able to stop the audio in the middle of the track.  While listening, users were required to complete a set of word puzzles, which simply entailed unscrambling letters to form phrases matching a provided clue, and were told to complete as many as possible before the audio terminated.  During this process, if participants heard a musical "signal", described as a musical anomaly that was attempting to draw their attention away from the task at hand, they were asked to click a button on the interface labeled, "I Heard a Signal!".  When the button was pressed, a text window on the interface either displayed a hint pertaining to the exam (for example, "Skip question 7, it is the hardest one!"), or displayed a neutral message ("No hint right now, keep going!").  This was done as an incentive for participants to remember to acknowledge perceived music modifications, as a means of encouraging a balance of attention between the artificial cognitive load task and the signal identification mechanism.

Upon completion of the audio stream, the response text fields for the word puzzles and the audio play button were rendered invalid, and participants were directed to a set of survey questions at the bottom of the interface.  The survey documented a fine-grained rating of their familiarity with the exact song and genre of the audio they had just heard, the amount of musical exposure they faced on a daily basis, and the primary contexts in which this exposure occured.  The survey is reprinted in Section \ref{appendix} as a supplementary reference.

The set of 10 tasks consisted of two well-known audio samples from each of the four genre categories (Classical, Blues, Jazz, and Pop), and two samples with tracks selected at random from the four categories which served as a control element in the study.  In each of the first 8 samples, 6 genre-specific modifications, with 2 at each level of subtlety, were added to the audio by the SoundSignaling system at random timestamps and ordinal positions.  For the last two samples, 6 samples of a control tone, a standard audio notification sample, was inserted at random timestamps and oridinal positions.  10 word puzzle sets consisting of 8 puzzles each were compiled, and randomly assigned to an audio task when given to a participant. Each puzzle set additionally contained a set of 5 hints, which were displayed at random when the "I Heard a Signal!" but was pressed, regardless of whether a modification had been correctly identified or not. The timestamps associated with the button clicks, referenced against the start of the audio track, were stored and correlated against the pre-determined timestamps and subtlety levels (with an acceptable click latency of 5 seconds from the end of the modification window) to form the analysis described below.

\subsection{Outcomes}

\begin{figure}
\includegraphics[width=0.45\textwidth]{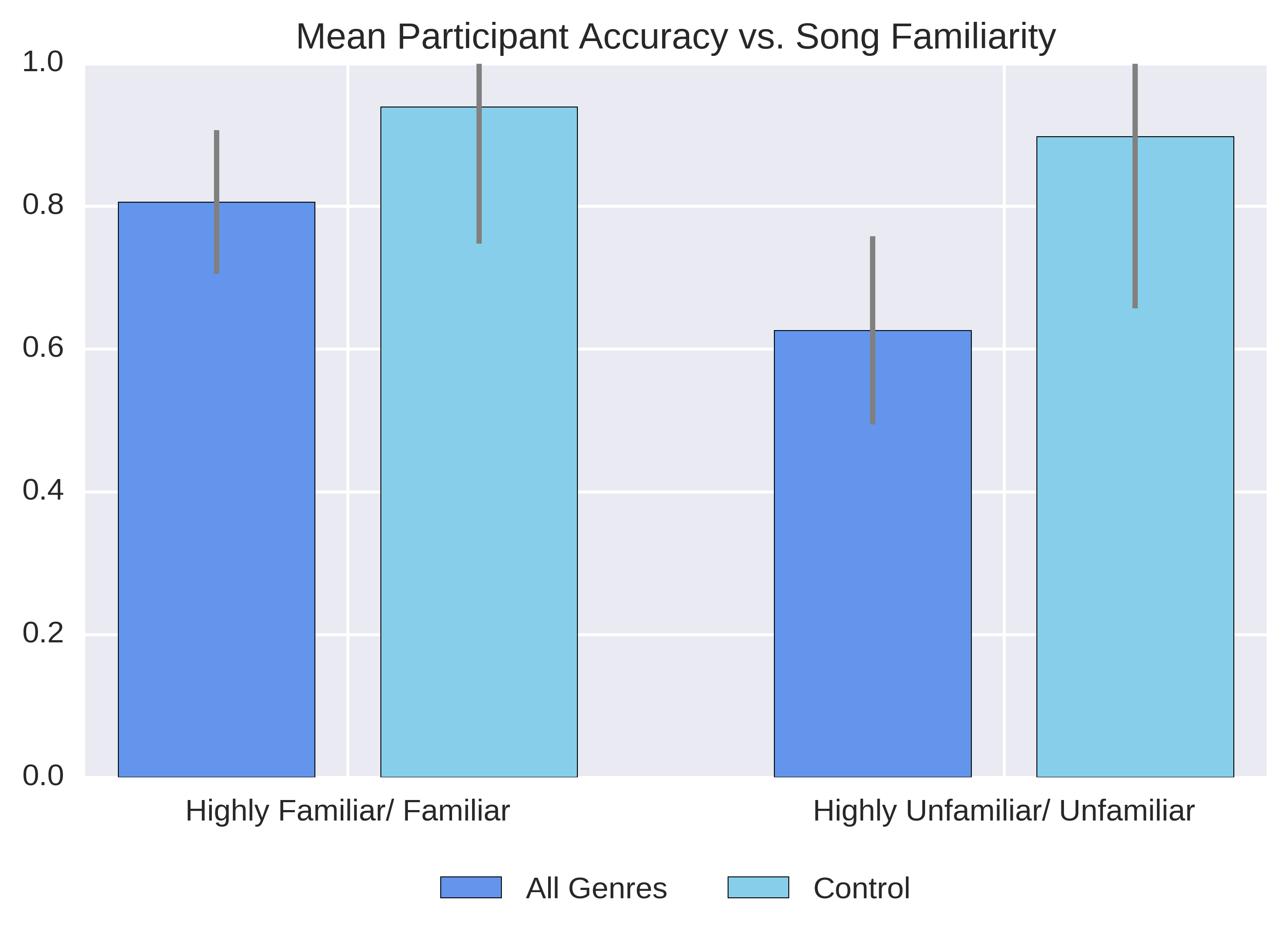}
\caption{Overall mean participant accuracy versus song familiarity, at a course response resolution; the mean for the control samples are shown in both groups as a reference.}
\label{Study1_Fig1}
\end{figure}

\begin{figure}
  \includegraphics[width=0.45\textwidth]{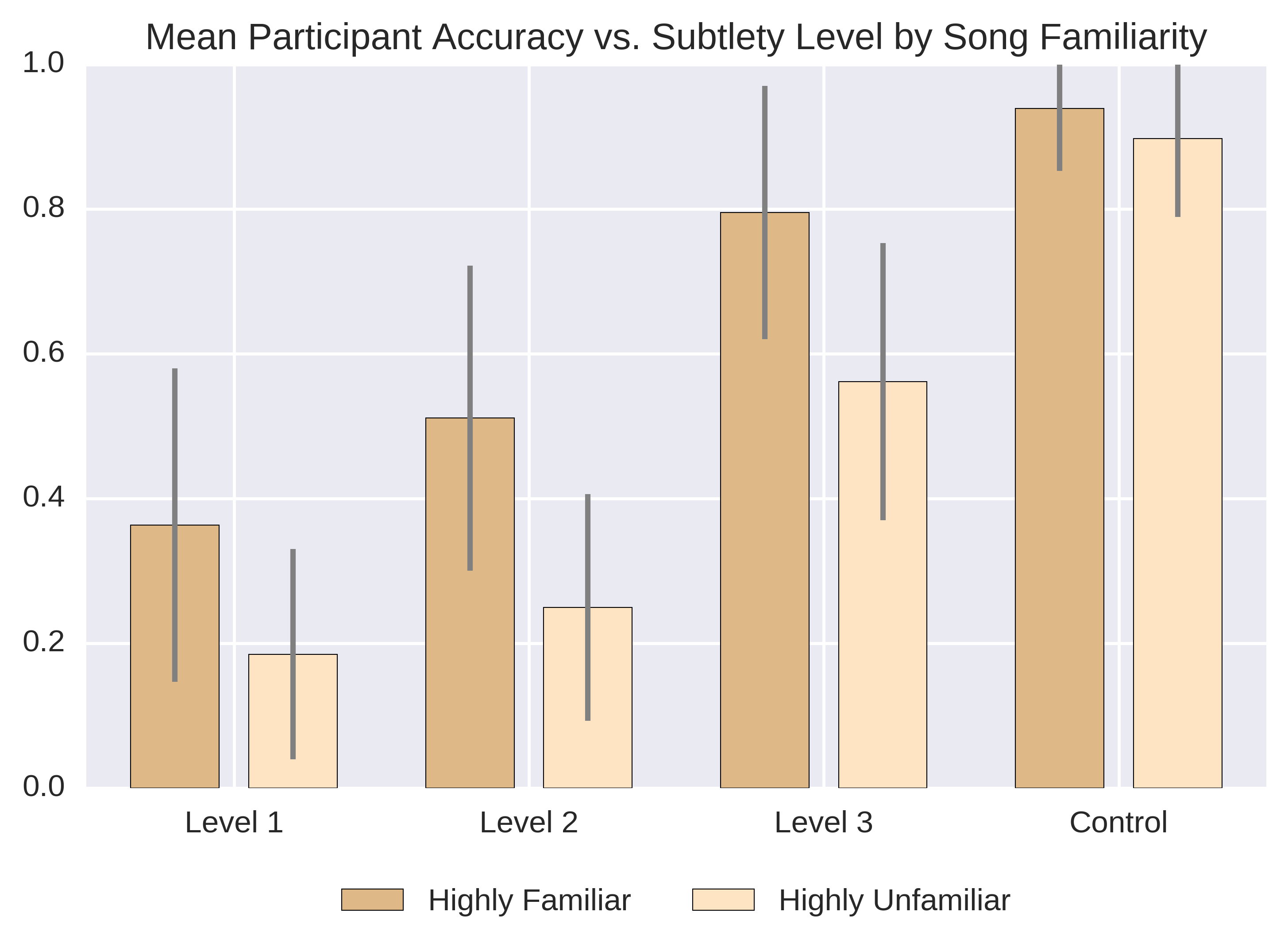}
  \includegraphics[width=0.45\textwidth]{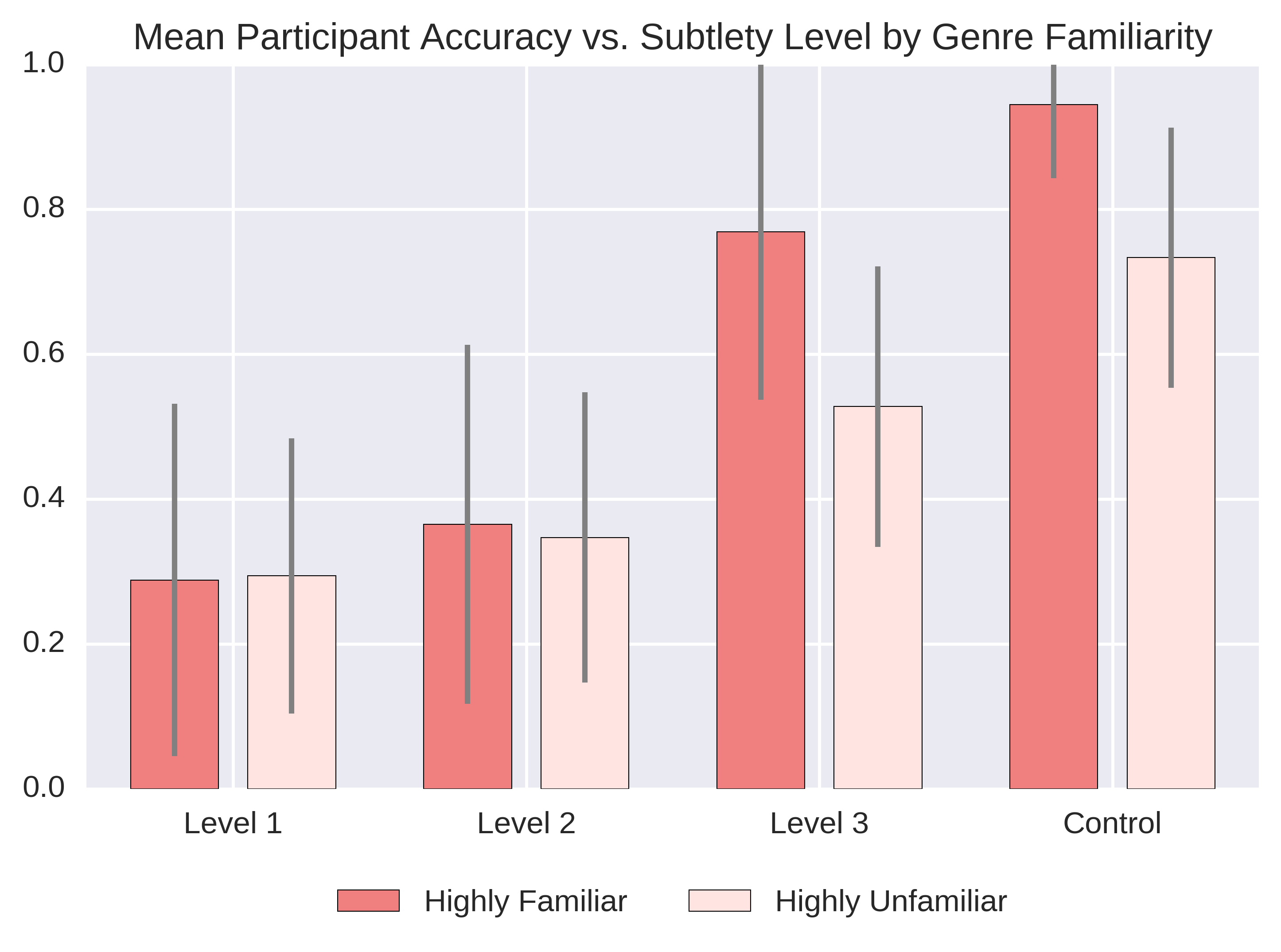}
  \caption{Left: Mean participant accuracy versus separate subtlety levels/ control, comparing tasks for which songs were rated "Highly Familiar" and "Highly Unfamiliar"; Right: Mean participant accuracy versus separate subtlety levels/ control, but comparing tasks for which song genres were rated "Highly Familiar" and "Highly Unfamiliar".}
  \label{Study1_Fig23}
\end{figure}

Firstly, the post-experiment survey data was analyzed for majority context regarding participant musical exposure.  It was ascertained from the data that approximately 57\% of the participant demographic listened to music for at least one hour a day, 67\% of participants had no formal training or performance experience in music, and 50\% of participants primarily listened to music as they worked. 

We then compute a high-level breakdown of the overall results, as shown in Figure \ref{Study1_Fig1}.  The plot shows a comparison of the mean per-participant accuracy in identifying modifications between two groups -- the group of participants who used the "Song Familiarity" field in the survey to identify the audio track as "Highly Familiar" or "Familiar" (left), and the group of participants who identified the audio as "Unfamiliar" or "Highly Unfamiliar" (right), agnostic to genre or subtlety level.  The mean accuracy results from the control samples in both groups are shown as a reference.  The increase in recognition rate from the second group to the first is shown to be statistically significant using an independant t-test, with p $<$ 0.05.  While this result is a first step in demonstrating the potential of the system, it is much more meaningful to consider the results \textit{only} from the set of tasks in which participants expressed being "Highly Familiar" with the specific track played, given the assumptions presented at the start of the work.  In Figure \ref{Study1_Fig23} (left), we show the mean per-participant identification accuracy broken down by subtlety level, and a comparison between task results from the "Highly Familiar" and "Highly Unfamiliar" participant samples (53 and 67 independent online tasks respectively, with an approximately even distribution of genres and subtlety levels/ control tones), agnostic to genre.  A fourth bar demonstrating the accuracy results from the control samples is given for reference.  The figure shows a monotonic, statistically significant (ind. t-test, p $<$ 0.05) increase in accuracy with an increase in subtlety level for the "Highly Familiar" category, and a statistically significant increase between the "Highly Familiar" and "Highly Unfamiliar" groups.  By contrast, we show results in Figure \ref{Study1_Fig23} (right), a comparison between genre familiarity/ unfamiliarity (58 and 53 independent online tasks respectively, with an approximately even distribution of genres and subtlety levels/ control tones) instead of song familiarity/ unfamiliarity, which does not express the trends to the same level of significance.  We believe this to be an important finding in support of our primary assumption -- that, by design, the utility of the system is driven by a user's intimate understanding of or familiarity with the music with which he/she may be using it. 

\begin{figure}
  \includegraphics[width=0.45\textwidth]{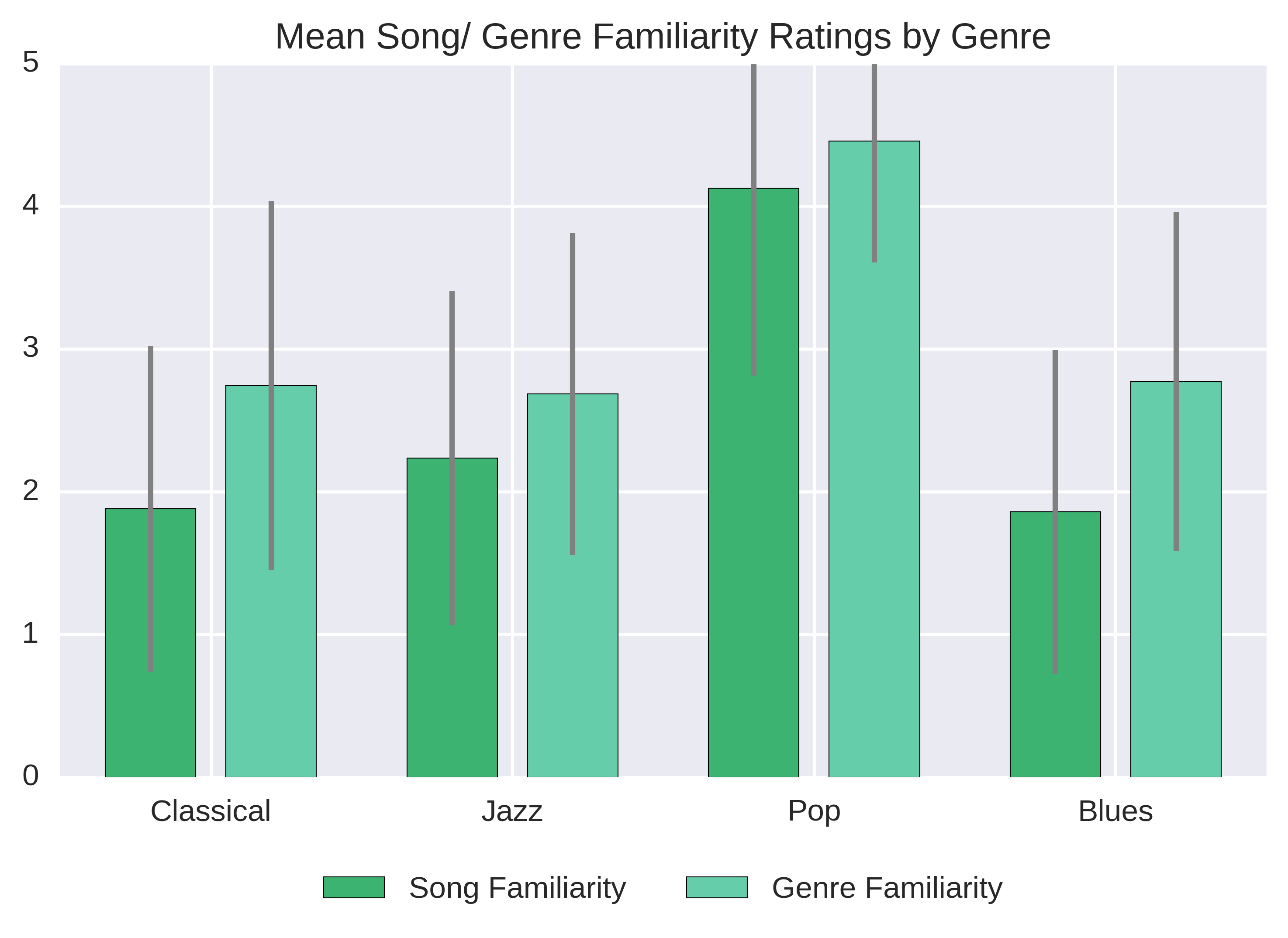}
  \includegraphics[width=0.45\textwidth]{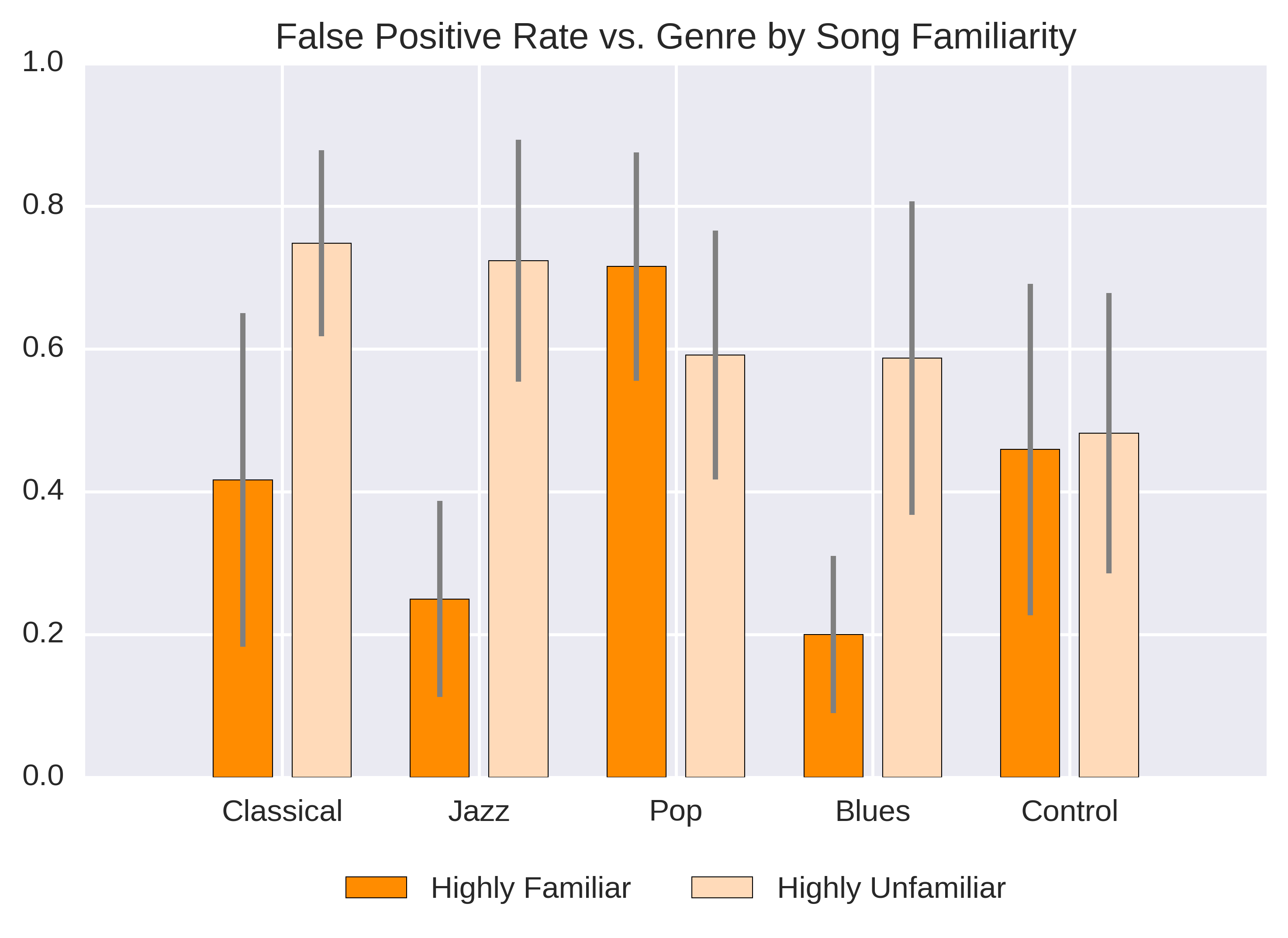}
  \caption{Left: The distribution of song familiarity and genre familiarity ratings, by genre. A score of "1" is "Highly unfamiliar", whereas a score of "5" is "Highly familiar". It has been inverted from the original questionnaire given in Section \ref{appendix} for readability. Right: The mean participant false positive rate, by genre.}
  \label{Study1_Fig45}
\end{figure}

It is important to consider these results in the context of the distribution of "Highly Familiar" songs, resulting from the online worker demographic. Figure \ref{Study1_Fig45} (left) shows the distribution of song familiarity and genre familiarity scores across all workers for each of the four genres.  The "Pop" genre rating distribution is skewed in favor of greater song and genre familiarity, and thus task samples in this genre are likely to have influenced the results more than the other genres, which demonstrate a more normal distribution.  This is expected from the very definition of "Popular" music, and highlights a specific area of focus for future design iterations of the modification system.

We next consider a breakdown of the false positive rate (defined as the number of false positives per total number of clicks per user, averaged across users) per genre, as shown in Figure \ref{Study1_Fig45} (right).  We again show a comparison of this quantity between users with the highest and lowest song familiarity ratings.  For the Classical, Jazz, and Blues genres, we demonstrate a statistically significant difference (ind. t-test, p $<$ 0.05).  We note, however, that this difference is absent for the Pop and Control categories.  We suspect that this is an artifact of the testing process -- in the instruction section of the web interface, participants were given minimal explanation regarding the nature of the "anomalies" they would be perceiving, which may have led participants to expect low-level audio feature manipulation (as was done for Jazz, Classical, and Blues) instead of/ in addition to audio scrambling or a default tone being inserted (the full text of the interface is reprinted in Section \ref{appendix}).  This behavior also naturally raises the question regarding the role of training and repeated experience with this system in affecting sensitivity to notifications, which is an additional area for further study.

We have demonstrated the technical functionality of the system by quantifying changes in identification performance as a function of familiarity and subtlety, as motivated in the start of the work.  However, quantifying the "softer" benefits of such a system in comparison to the control is a more difficult challenge, one that is beyond the scope of this work in a complete sense, primarily limited by the infeasibility of running controlled, large-scale experiments with highly personalized music choices.  However, to collect anecdotal evidence that may suggest these benefits, that demonstrate the usage of the system in a real world context affording user-defined musical input, and that open avenues for further user-focused research, we consider a discussion from a small-scale, in-the-wild experiment described in the section below.

\section{In-the-wild Evaluation: Anecdotal Evidence}\label{inthewild}

\subsection{Goals}

In order to understand the behaviors and preferences of a small group of users engaged with a custom system for an extended period of time, we choose to perform a small-scale, longer-term experiment.  The primary motivation was to examine the system when used with tracks highly personalized for individual users, by allowing them to select their own tracks.  In a strictly qualitative sense, we sought to gain insight into whether users were likely to consistently use the system over an extended period of time, and if so, how they perceived the algorithmic choices for modifications from a musical perspective, given personalized music selections.

\subsection{Study Procedure}

For our in-the-wild evaluation, study participants were provided with a custom installation of the SoundSignaling system as a command line tool on a personal computer, and were instructed to prepare a locally accessible collection of music with at least 50-100 tracks per participant in advance of the start date.  The version of the software provided to the participants included a sample client application designed to monitor a GMail account using Google's GMail API \cite{gmailAPI}, by providing a call through the SoundSignaling Client API if new emails had entered a user's primary inbox since the last time the client application had checked. We chose email monitoring as a primary client task for the purpose of the study, since it has been established that email notifications contribute signifcantly to the average number of notifications received by our target population and will be broadly applicable as a monitoring need \cite{pielot2014situ, mehrotra2015designing}.

Participants were instructed to use the software system over the course of a 10-day period, primarily when they would normally be listening to music while completing an additional task.  After individual meetings to assist with installation and setup, participants were asked to complete a pre-study interview before beginning to use the system (see Section \ref{appendix}).  Over the course of the trial period, participants were told that they could fine tune the frequency at which their email was checked as well as the level at which the modification was made, by means of command line arguments everytime the system was started for music playback.  Finally, after the 10-day trial period, participants were asked to complete a post-study interview (see Section \ref{appendix}), as well as to upload a log file produced by their system.  The log file consisted of system start timestamps, stop timestamps, signal level values, notification-receipt timestamps, and responses to a system prompt asking for a description of the task being engaged in while listening to music.

We originally recruited 13 student participants at random from our university (none of whom were affiliated with our research group), under the condition that a participant must be willing to participate for 10 days, that he/ she uses GMail as his/ her primary email client, and is comfortable with using a command-line application. However, the behaviours and qualitative findings from only 6 participants (4 male and 2 female) are presented here, after filtering for inadequate or incorrect use of the system.  While the long-term usage and 1-to-1 interaction required to help each participant with installation and usage constrained the study to a small set of participants, we believe the qualitative findings discussed below are meaningful in assessing the current work and identifying areas of focus for future research.  

\subsection{Outcomes}

Over the course of the trial period, the six participants collectively listened to more than 300 tracks, used the system for 180 hours and 67 independent listening sessions, and received 157 email notifications while engaged in a broad variety of additional activities such as "surfing the internet", "coding on the couch", "browsing social media and playing games", "answering a homework about signal processing", and "catching up on emails". Additionally, 4 of 6 participants rated their taste in music and playlist content across listening sessions to be constrained to a few genres and fairly narrow (see Section \ref{appendix}), while the remaining two exhibited substantial diversity in both categories.  All participants mentioned that they listened to music for 1 - 5 hours each day.  We present our findings summarized from the qualitative opinions collected from these participants through both the post-study survey and post-study interviews below, and have taken the liberty of grouping the responses into broad themes relevant to the introductory motivation.

\textit{\textbf{Algorithmic Choices and Musicality.}} Speaking to the musicality of the modifications, there was unanimous agreement from the participants about the effectiveness of the genre-specific processing and musical relevance:
\begin{quote}
"I liked how well it integrated into the music I was already listening to, especially with the section-switching in the pop songs. The notifications felt very musically satisfying."
\end{quote}
\begin{quote}
"I really enjoy the musicality of the system and the seamlessness of it."
\end{quote}
\begin{quote}
"Really impressed with how well this matched some of my music!"
\end{quote}
\begin{quote}
"It made me look forward to receiving mail as I would be 'rewarded' with a passage of musical interest!"
\end{quote}

\textit{\textbf{Switch Cost.}} Even though users were not specifically prompted to address the notion of a "switch cost" during interviews, many expressed a greater sense of control over responding to notifications, the elimination of an additional information-carrying modality, and subtle delivery of content.  Select quotes representative of these findings are below:
\begin{quote}
"I like that I don't have to physically open my email tab to know that I have new email coming in; it helps me check my email less."
\end{quote}
\begin{quote}
"I can receive notifications about my email without having to look at my phone screen for notifications or open up a Gmail tab."
\end{quote}
\begin{quote}
"I think I'd prefer not to be actively notified [of email] at all, but if I had to, I'd prefer the subtlety of this!"
\end{quote}

\textit{\textbf{Cognitive Load and Missed Notifications.}} Amongst the most important findings were the ones that addressed participants' interaction with and perception of the system when they mentioned using it during intervals of heightened cognitive load or focus, such as when completing assignments or drafting important emails.  For example, one participant stated:
\begin{quote}
"Actually, I may even prefer it [the system] to no notifications -- I didn't hear many of the notifications while I worked (a plus), and I noticed a couple when I wasn't focused and [was] paying more attention to the music."
\end{quote}
Many users similarly expressed the notion that they were less likely to perceive the musical notifications when under increased cognitive load, suggesting that this work provides a natural, passive index into a user's cognitive load, without the need for computationally intensive activity or behavior patternization.  However, in order for the system to exhibit this principle, participants noted that they would have liked some means by which to retrieve those missed notifications at a later time.  For example, one participant mentioned:
\begin{quote}
"I'm not confident that I hear all of the notifications, especially when focused.  I think this is again largely a pro during concentration time, however it still leaves me with an ambient unease that I missed an email and I need to compulsively check my phone/email when I think of it."
\end{quote}
This highlights what is perhaps one of the biggest challenges stemming from the SoundSignaling system -- the possibility of missing notifications in a scenario when one \textit{would} like to receive them (such as when listening to music while under fairly minimal cognitive load).  In response to this, participants themselves collectively identified several features that could form effective solutions: tunable feedback mechanisms, embedded learning to optimize for user recognition of notifications, simple notification summaries that could be used manually by a user as a means of training themselves, or priming themselves to the types of audio modifications they might (or might not) perceive.  One participant suggested:
\begin{quote}
"I'd also love a self-tuning thing, something that blinks in the corner 10 seconds after a notification so you get trained to recognize the types of notifications and can keep mental track of how many of these notifications you actually hear/miss.."
\end{quote}
As a different solution, another participant suggested:
\begin{quote}
"Having some feedback mechanism to ensure the notification is noticed eventually would be nice.  I wonder about starting with less obvious notifications and slowly making them more obvious with a very slow cadence, or learning when someone checks their email and find[ing] a notification intensity that works 25\% of the time and repeating it slowly until the user responds."
\end{quote}
These suggestions form an excellent roadmap for future user-focused studies in attempting to better understand and address this problem.

\textit{\textbf{A New Paradigm in Music Listening.}}  Finally, many of the participants expressed noticing a conscious change in the way they listened to their music over the course of the trial period. One participant said:
\begin{quote}
"I found that I paid more attention to transitions and odd compositional choices in the music I'm familiar with than I normally do, which is cool."
\end{quote}
\begin{quote}
"[I was] listening more intently and questioning whether choices/reverbs/weird effects and transitions were always there and chosen by the artist or added afterwards."
\end{quote}
However, several participants noted that this resulted in some ambiguity. For example:
\begin{quote}
"It [the system] changes the way I listen to my music.  I'm more attentive to small details and more analytical, and there is some confusion.  In a lot of cases, I really enjoyed this change- it's like a bit of a game and it draws you into focused analysis of the songwriting.  But over more time I might not want to have that feeling, or be unsure whether I have an email or whether the songwriter made an interesting choice."
\end{quote}

This raises a fundamental point that came to light as a result of the study: use of the system might elicit, at least as an initial by-product, a prolonged attentional shift towards the background audio as users are intrigued by the novelty of the modifications and look forward to observing them.  However, in an even more long-term characterization, it would be important to understand whether this effect would wear off over time as users become more capable of patternizing the nature of the modifications, or whether there will continue to be a "start-up" effect associated with playlists or styles of music that are new to a user.  Moreover, it would be of interest to comprehend the impact of this phenomenon on the adoption and retention rate of such a technology.

\section{Limitations and Future Work}

 From a technical standpoint, the majority of the critical feedback provided by the participants pertained to the user interface.  While the design of an intuitive and robust UI was outside the scope of this work, development in this area will allow us to share the platform more broadly and collect data over several months.  Currently, the command line tools allow for songs uploaded to the server to be played strictly in the order of preprocessing, with only a start and stop functionality. Specifically, participants requested features such as pause functionality, playlist shuffling, and on-the-fly playlist reorganization.  We intend to add support for this functionality, in addition to a basic graphical or browser-based interface to replace the current set of command-line tools.

Moreover, the software application as it stands is fairly computationally intensive.  Due to the range of features being collected for every song in addition to real-time genre classification,  pre-processing requires as much time as 10-15\% of the duration of the song.  This deterred some of the study participants from utilizing the platform more excessively.  In future work, we will consider an alternate, light-weight implementation to understand if similar, music compatible modifications could be generated with substantially less pre-processing.

Lastly, users pointed out that the need for locally accessible music was often a significant barrier.  Participants repeatedly asked if the platform couldn't be "plugged in to Spotify", and took several days to collect a library of music if they didn't have one readily available.  This is an important trade-off to consider -- arguably, the "seamless", musically relevant modfications were able to generated as a result of the fine-grained pre-processing on each and every track, and the ability to pre-process is lost with use of a streaming service.  As mentioned above, we intend to develop a version of this system that does not require pre-processing, but produces real-time musical "effects" after the online estimation of certain parameters (such as tempo and amplitude).  We intend for this online estimation to be the result of an embedded learning model, pre-trained on a large corpus of songs.

From a user behavior standpoint, one limitation brought to our attention by users themselves was the possibility of missing notifications in contexts where one might not ordinarily mind being disturbed or distracted; while it is true that notifications may always be retrieved in a traditional manner (i.e., checking one's phone every few hours despite not having heard an SMS alert tone), participants noted the absence of mechanisms for training or more fine-grained control over subtlety settings to minimize the need for this (see Section \ref{inthewild}).  We consider the idea that one's ability to recognize a notification is driven by two parammeters -- one's intimate familiarity with the musical track, as suggested by our crowd sourced results, and one's understanding of "what to look for", or expectations of the nature of the notification.  As suggested by participants through the anecdotal evidence, training mechanisms that provide feedback through another medium (visual, for example), could be implemented to help user develop their recognition abilities; however, unlike traditional audio sonification platforms, the system was intentionally designed to avoid explicit training phases where mappings between audio samples and cues are learned in advance. This discussion point suggests a careful study of the tradeoff between aesthetically-motivated generalization, training mechanisms, and missed notifications, especially in the context of music that may be less familiar to a user (such as a well-known genre but a new artist).

Finally, we underscore the point that this system treats notifications as a fluid concept with a dynamic embodiment, in a way that is somewhat orthagonal to the definition of a notification. While literature suggests that this treatment may provide benefits in terms of attention (though significant further study would be required to show this), we consider the prospect of more wholistic changes in user behavior as individuals attempt to receive information in this non-traditional manner.  Amongst other things, notifications delivered in this way combine two sources of stimulus into one, affecting our perception of both channels; our notion of the "reliability" of notifications is altered, and must be taken into account when mapping sources to subtlety levels or to decide whether to use the system at all (using the system to provide a morning wakeup alarm, for example, is probably not recommended!); our attentiveness towards the audio may vary as a function of time and experience with the system, as mentioned above, from actively "searching" for notifications to being passively alerted of them; and finally, we may be subconciously provided with a model of our cognitive activity as a function of missed/ detected notifications, resulting in a heightened degree of self-awareness.  We posit that, once the technological challenges mentioned above have been surmounted, it is critical to study each of these phenomena in detail with future broad scale experiments.

\section{Conclusion}
In this work, we have demonstrated the SoundSignaling platform, a novel method for delivering real-time notifications via stylistic, genre-specific music modifications.  Using a design inspired by the notion of cognitive incongruence and informed by musical principles, we first validate the system's algorithmic components and operational choices through a crowd-sourced study, and qualitatively discuss further implications pertaining to switch cost, user cognitive load, and listening behavior by considering anecdotal outcomes from a small-scale, in-the-wild usage experiment.  Finally, we highlight several areas for further exploration, including training and feedback mechanisms, subtlety levels that are fine-tuned by play history, and online models for operating on streaming music. Ultimately, the results suggest that this work may the first step towards re-thinking the age-old paradigm of binary audio notifications using a relatively unexplored, now ubiquitous audio landscape -- a personal collection of music.

\appendix\label{appendix}
\section{Crowd-Sourced Evaluation: Complete Interface}

\begin{figure}[H]
  \includegraphics[width=1.0\textwidth]{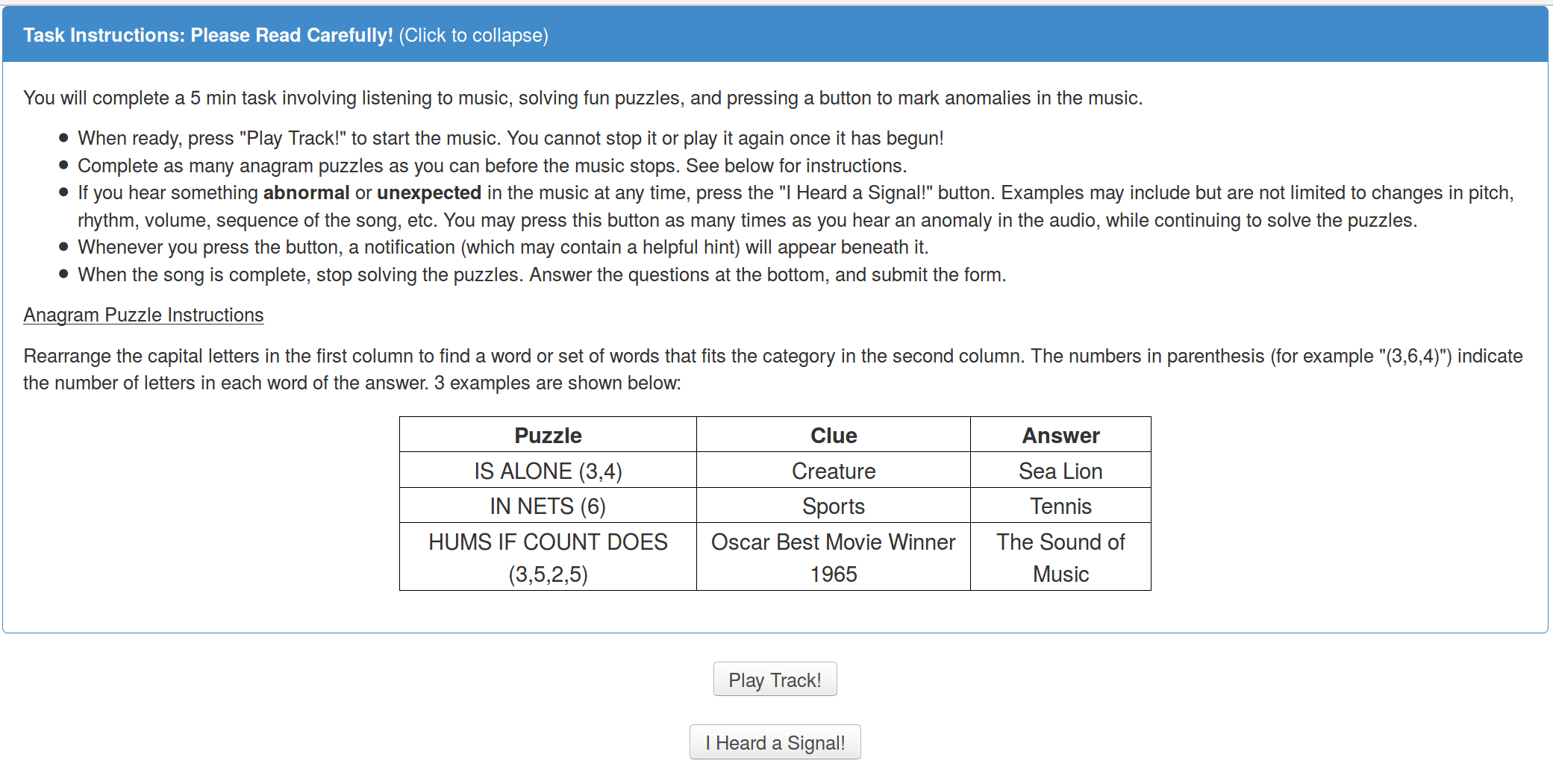}
\end{figure}
\begin{figure}
  \includegraphics[width=1.0\textwidth]{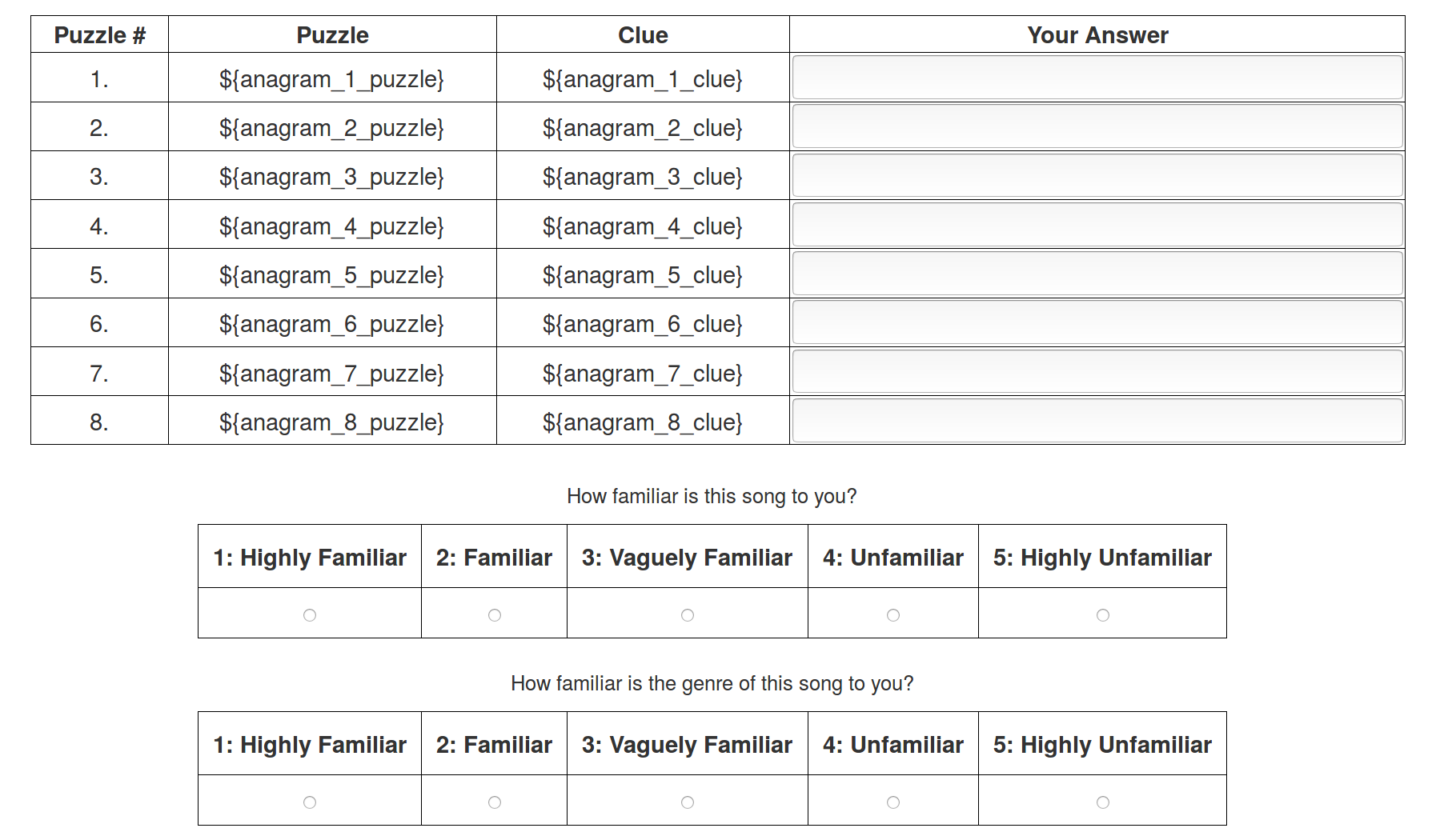}
\end{figure}
\begin{figure}
  \includegraphics[width=1.0\textwidth]{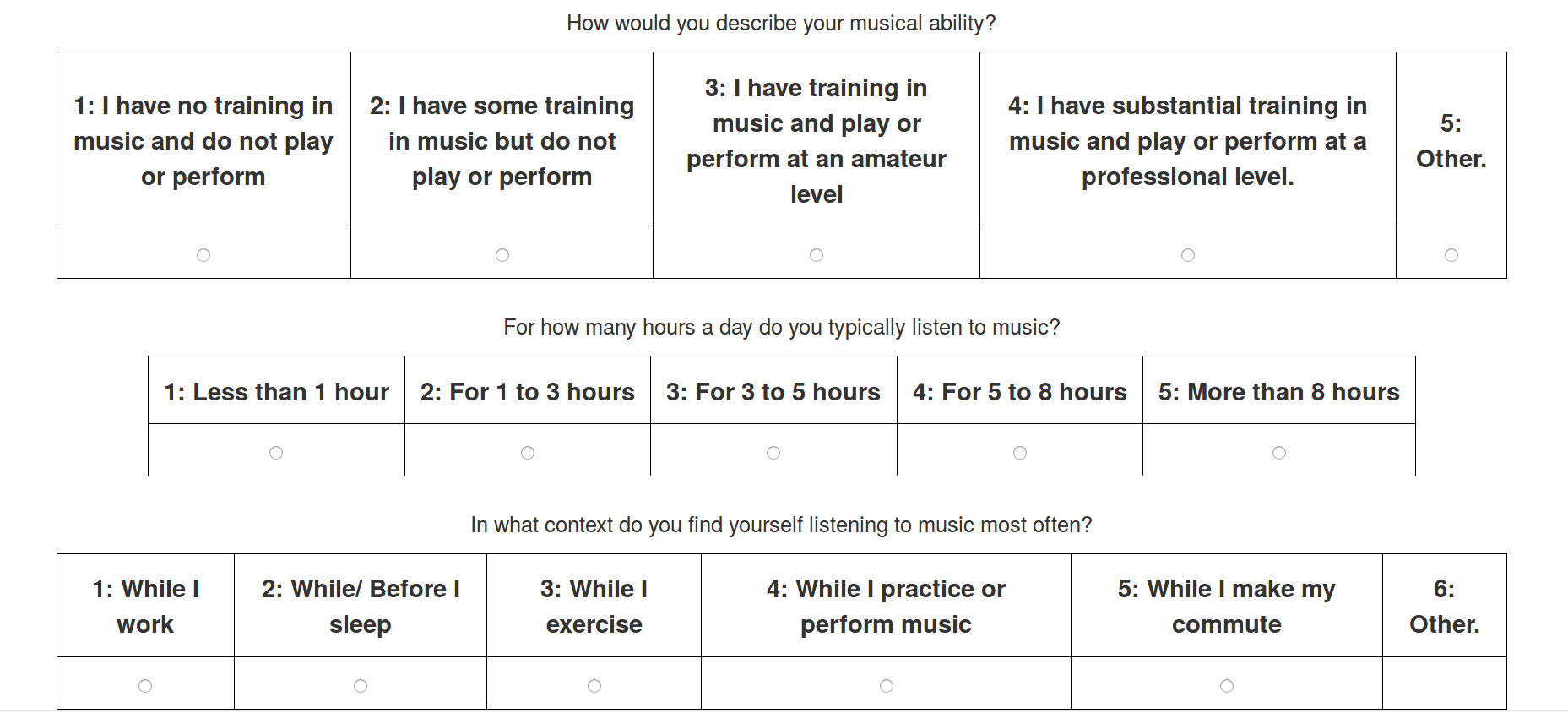}
\end{figure}

\clearpage

\section{In-the-wild Study: Pre-Study Questionnaire}
\begin{enumerate}
  \item How do you currently prefer to receive notifications about incoming email?
  \begin{enumerate}
    \item Active visual notification (Ex: a pop-up)
    \item Passive visual notification (Ex: a number increasing on your gmail tab)
    \item Audio notification (Ex: a short ring or beep)
    \item Haptic notification (Ex: a phone vibrating)
  \end{enumerate}
  \item On a normal day, how many times a day would you estimate that you check your email?
  \item Speaking for yourself, would you say that this is ideal, too often, or too infrequent?
  \begin{enumerate}
    \item Ideal
    \item Too often, I really should be checking it less)
    \item Too infrequent, I really should be checking it more
  \end{enumerate}
  \item To what extent do you feel that these notifications distract you from the task at hand? \newline
  \textbf{(I find them to be extremely distracting)}  1 | 2 | 3 | 4 | 5 \textbf{(I don't find them to be distracting at all)}

  \item How likely are you to respond to notification when you perceive one? \newline
  \textbf{(Very unlikely -- I typically ignore it until a later time)}  1 | 2 | 3 | 4 | 5 \textbf{(Very likely -- I typical respond immediately)}

  \item If I receive a notification when I am engaged in a task that requires more mental effort than my usual work, ..
  \begin{enumerate}
    \item I am more likely to ignore the notification
    \item I am more likely to respond to the notification
    \item I respond as usual, independent of the nature of the task
  \end{enumerate}

  \item For how many hours a day do you typically listen to music?
  \begin{enumerate}
    \item Less than 1 hour
    \item For 1 to 3 hours
    \item For 3 to 5 hours
    \item For 5 to 8 hours
    \item More than 8 hours
    \item Other:
  \end{enumerate}

  \item Which of the following best applies to you?
  \begin{enumerate}
    \item I find that listening to music typically impedes productivity of the task I'm working on
    \item I find that listening to music typically boosts productivity of the task I'm working on
    \item I find that listening to music typically has no effect on the productivity of the task I'm working on
  \end{enumerate}

  \item How would you rate the diversity of your taste in music? \newline
  \textbf{(I listen to a very diverse set of genres and musical styles)}  1 | 2 | 3 | 4 | 5 \textbf{(I typically listen to a narrow set of genres and musical styles)}

  \item How would you rate the diversity of your playlists or collections of songs across listening sessions? \newline
  \textbf{(My playlists are constantly changing -- I shuffle through large collections of music)}  1 | 2 | 3 | 4 | 5 \textbf{(My playlists are pretty static -- I listen to the same songs for several listening sessions at a time)}

\end{enumerate}

\section{In-the-wild Study: Post-Study Questionnaire}

\begin{enumerate}
  \item Overall, did you prefer this system over your usual means of receiving email notifications? Why or why not?
  \item What, specifically, did you like about this system?
  \item What, specifically, did you dislike about this system?
  \item While using the system over the course of the week, how many times a day would you estimate that you checked your email?
  \item Speaking for yourself, would you say that this is ideal, too often, or too infrequent?
  \begin{enumerate}
    \item Ideal
    \item Too often, I really should be checking it less)
    \item Too infrequent, I really should be checking it more
  \end{enumerate}

  \item To what extent do you feel that these alternative notifications distracted you from the task at hand? \newline
  \textbf{(I found them to be extremely distracting)}  1 | 2 | 3 | 4 | 5 \textbf{(I didn't find them to be distracting at all)}

  \item When using the system, how likely were you to respond to notification when you perceived one? \newline
  \textbf{(Very unlikely -- I typically ignored it until a later time)}  1 | 2 | 3 | 4 | 5 \textbf{(Very likely -- I typical responded immediately)}

  \item Which of the following best applies to you?
  \begin{enumerate}
    \item I found that using the system typically impeded productivity of the task I was working on
    \item I found that using the system typically boosted productivity of the task I was working on
    \item I found that using the system typically had no effect on the productivity of the task I was working on
  \end{enumerate}

  \item What, if any, parameters did you modify or fine tune to your liking? Why? (Ex: Email check frequency, obviousness level, etc)

  \item If I received a notification through the system when I was engaged in a task that required more mental effort than my usual work, ..
  \begin{enumerate}
    \item I was more likely to ignore the notification
    \item I was more likely to respond to the notification
    \item I responded as usual, independent of the nature of the task
  \end{enumerate}

  \item If such a system were available, would you be likely to use it on a frequent basis? Why or why not?

  \item What improvements would you like to see in a deployed version of this tool?

  \item Please upload your log file. 

\end{enumerate}

\begin{acks}

The authors would like to thank David Ramsay, Spencer Russell, and Gershon Dublon for their invaluable technical feedback; Artem Dementyev and Nan Zhou for their feedback regarding the publication; and the study participants for their time and dedication.

\end{acks}

\bibliographystyle{ACM-Reference-Format}
\bibliography{signaling}

\end{document}